\documentclass[conference,12pt,onecolumn]{IEEEtran}
\usepackage[T1]{fontenc}
\usepackage{cite}
\usepackage{amsmath}	
\usepackage{graphicx}
\usepackage{epstopdf}
\usepackage{subcaption}
\usepackage{algpseudocode,algorithm,algorithmicx}
\usepackage{amssymb}
\usepackage{float}

\DeclareMathOperator{\atantwo}{atan2}

\renewcommand{\vec}[1]{\mathbf{#1}}	

\DeclareMathOperator*{\argmax}{arg\,max}

\ifCLASSINFOpdf
\else
  
\fi

\begin{document}
%
\title{Joint Localization and Mapping through Millimeter Wave MIMO in 5G Systems}


\author{\IEEEauthorblockN{Rico Mendrzik\IEEEauthorrefmark{1}, Henk Wymeersch\IEEEauthorrefmark{2}, Gerhard Bauch\IEEEauthorrefmark{1}}\IEEEauthorblockA{\IEEEauthorrefmark{1}Institute of Communications,
Hamburg University of Technology, Hamburg 21073, Germany}
\IEEEauthorblockA{\IEEEauthorrefmark{2}Department of Electrical Engineering, Chalmers University, Gothenburg 412 58, Sweden}}

%


\maketitle

\begin{abstract}
Millimeter wave signals with multiple transmit and receive antennas are considered as enabling technology for enhanced mobile broadband services in 5G systems. While this combination is mainly associated with achieving high data rates, it also offers huge potential for radio-based positioning. Recent studies showed that millimeter wave signals with multiple transmit and receive antennas are capable of jointly estimating the position and orientation of a mobile terminal while mapping the radio environment simultaneously. To this end, we present a message passing-based estimator which jointly estimates the position and orientation of the mobile terminal, as well as the location of reflectors or scatterers in the absence of the LOS path. We provide numerical examples showing that our estimator can provide considerably higher estimation accuracy compared to a state-of-the-art estimator. Our examples demonstrate that our message passing-based estimator neither requires the presence of a line-of-sight path nor prior knowledge regarding any of the parameters to be estimated. 
\end{abstract}

\section{Introduction}
\subsection{Motivation and State of the Art}
In many conventional wireless networks, accurate radio-based positioning relies on the existence of a line-of-sight (LOS) path between the transmitter and the receiver. Based on the signaling and antenna apertures, position-related parameters can be derived from the received signal. Such parameters include the time-of-arrival (TOA), angle-of-arrival (AOA), angle-of-departure (AOD), and received signal strength (RSS). Based on the capabilities of the systems, one or more of these parameters can be determined and leveraged for position estimation. For instance, lateration uses the TOAs with respect to multiple transmitters in order to obtain an estimate of the position of the receiver \cite{CTXC2004}, while angulation employs the AOAs with respect to multiple transmitters to estimate the position of the receiver \cite{G1996}. In contrast to many conventional systems, the millimeter wave (mmWave) multiple input multiple output (MIMO) physical (PHY) layer proposal in 5G enables the determination of a triplet of position-related parameters of every received multipath component: Due to the high temporal and spatial resolution of mmWave MIMO, the TOA, AOD, and AOA of every multipath component can be estimated \cite{SGDSW2015,SGDSW2017,SZASW2017}. Therefore, every non-line-of-sight (NLOS) path can be leveraged for position and orientation estimation \cite{MWBA2017}.  Even in the absence of LOS\footnote{We refer to the scenario where only NLOS component is as obstructed line-of-sight (OLOS).}, accurate positioning using only a single transmitter becomes possible if at least three NLOS paths exist \cite{MWBA2017}. Note that harnessing NLOS paths for position estimation clearly marks a paradigm shift in the field of radio-based positioning, where NLOS paths were conventionally considered as useless if no prior information is available \cite{WWS2017}. 

Recently, different estimators have been presented in the literature which employ NLOS paths for position estimation and mapping\cite{SGDSW2017,TVDW2017,URGW2016,GPUJZ2017}. In \cite{SGDSW2017}, a least-squares (LS) approach with extended invariance principle (EXIP) is used to recover the position and orientation of the receiver from the TOAs, AODs, and AOAs. This approach can be used in the presence and absence of LOS. However in the absence of LOS, the approach requires to solve a large number of parallel least-squares  problems. In particular, a fine-grained grid of trial orientations is created and one LS problem has to be solved for every trial orientation. The residuals of all solved problems are cached and only the solution with the lowest overall residual is retained. The drawback of this approach is that generally a fine granularity of the trial values for the orientation is required to achieve accurate estimates. In \cite{TVDW2017}, a Gibbs sampling-based approach is presented where an iterative sampling process is executed. The Gibbs sampler starts with an initial guess regarding the position and orientation of the mobile. Based on this guess, the positions of the reflectors or scatterers are determined and the initial guess on the position and orientation are updated sequentially. This procedure is repeated numerous times. A selection of all samples is retained and used for position and orientation estimation of the mobile, as well as for the estimation of the reflectors or scatterers. However, the authors in \cite{TVDW2017} did not show that their proposed Gibbs sampler works in the case of OLOS, i.e. when the LOS component is missing. In \cite{URGW2016,GPUJZ2017}, a sequence of observations including path delays and motion data is used to sequentially estimate the position of a mobile terminal and map the radio environment. This simultaneous localization and mapping (SLAM) approach requires multiple observations at different time instances. 
\subsection{Contribution and Paper Organization}
We present a novel message passing-based estimator that uses the concept of nonparametric belief propagation to determine estimates on the position and orientation of the mobile terminal, as well as estimates on the locations of reflectors or scatterers. We show that our message passing-based estimator provides accurate estimates in the OLOS scenario. Our main contributions are summarized as follows:
\begin{itemize}
	\item We propose a novel message-passing based estimator that jointly estimates the position and orientation of a mobile terminal along with the locations of the scatterers or reflectors in the case of OLOS without assuming any prior knowledge. 
	\item The proposed estimator is capable of performing accurate single-snapshot\footnote{By \textit{single-snapshot} SLAM we mean that the information from a single transmission burst is sufficient to estimate the location and orientation of the mobile terminal and to create a map of the radio environment. } SLAM even in the absence of the LOS path.
	\item Our numerical examples show that the root-mean-square error (RMSE) of the proposed estimator is often lower compared to the least-squares approach from \cite{SGDSW2017}.
\end{itemize}
The rest of the paper is organized as follows. Section \ref{sec:sys_model} discusses our system model. In section \ref{sec:BP}, we review the theory regarding our novel message passing-based estimator, while section \ref{sec:NBP} describes the particle-based implementation of the estimator. Section \ref{sec:numerical_example} contains numerical examples and section \ref{sec:conclusion} concludes the paper.

\section{System Model}
\label{sec:sys_model}
\subsection{System Model}
\begin{figure}[t]%
\centering
\includegraphics[width=0.98\columnwidth]{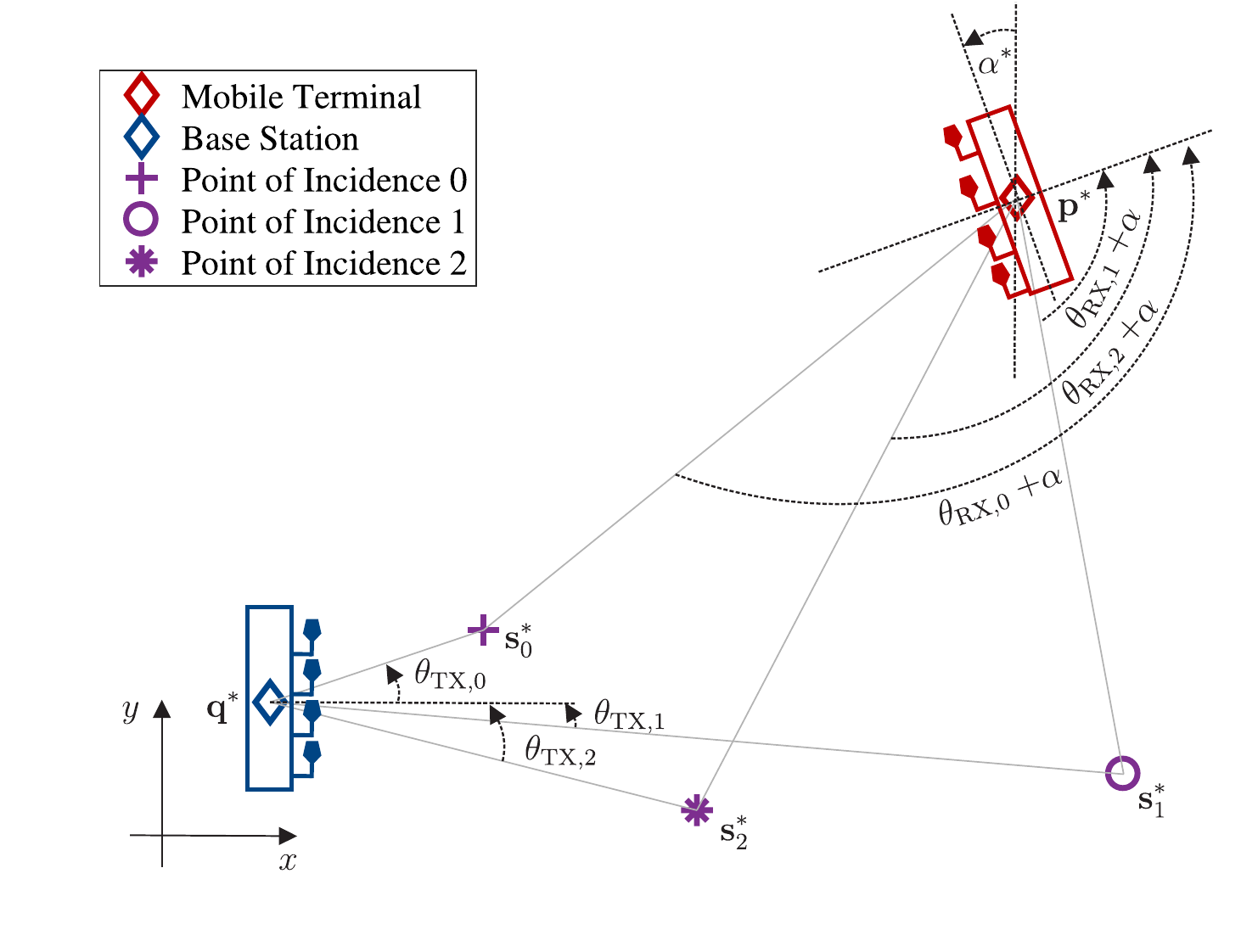}%
\caption{\textit{Geometry of the scenario -} A mobile terminal attempts to determine its unknown position $\vec{p}$ and orientation $\alpha$ using distance ($\hat{d}_j$), AOA ($\hat{\theta}_{\mathrm{RX,j}}$) and AOD ($\hat{\theta}_{\mathrm{TX,j}}$) measurements from multiple NLOS paths. Simultaneously, the mobile terminal estimates the locations of the points of incidence $\vec{s}_j$ corresponding to the NLOS paths.}%
\label{fig:geometry}%
\end{figure}
Fig. \ref{fig:geometry} depicts a scenario with $J=3$ NLOS paths. We consider a base station (transmitter) and a mobile terminal (receiver). The base station is located at the position $\vec{q}^*=[q_{\mathrm{x}}^*,q_{\mathrm{y}}^*]^{\mathrm{T}}$, while the mobile terminal is at $\vec{p}^*=[p_{\mathrm{x}}^*,p_{\mathrm{y}}^*]^{\mathrm{T}}$. The position and orientation of the base station are perfectly determined and known to the mobile terminal. Without loss of generality, we assume that the base station is at the origin and its array is aligned with the y-axis. The received signal comprises $J\geq 3$ NLOS components with associated points of incidence\footnote{Note that scatterers are objects that are much smaller than the wavelength of the signal, while reflectors are objects with a specific reflection point that are much larger than the wavelength of the signal. In order to cover both reflectors and scatterers, we use the term \textit{point of incidence }in place of the location of a scatterer and the point of reflection of a reflector.} $\vec{s}_j^*=[s_{\mathrm{x},j}^*,s_{\mathrm{y},j}^*]^{\mathrm{T}},\forall j$. Generally, the number of NLOS components in the mmWave band is small \cite{PK2011}.  In addition, due to the high path loss in the mmWave band, NLOS components are assumed to originate from single bounce scattering or reflection only \cite{HSZWM2016,GGD2017,DS2014}.  We assume that the receiver determines a triplet of estimates (TOA, AOD, and AOA) for every path as described in, e.g., \cite{SGDSW2017}. We refer to these estimates as observations and collect them in the vector
\begin{equation}
\hat{\vec{z}}=[\hat{d}_0,\hat{\theta}_{\mathrm{TX},0},\hat{\theta}_{\mathrm{RX},0},...,\hat{d}_{J-1},\hat{\theta}_{\mathrm{TX},J-1},\hat{\theta}_{\mathrm{RX},J-1}]^{\mathrm{T}},
\label{eq:observation}
\end{equation}
where $\hat{d}_j$, $\hat{\theta}_{\mathrm{TX},j}$, and $\hat{\theta}_{\mathrm{RX},j}$ denote the estimates on the distance, AOD, and AOA of the $j^{\text{th}}$ path, respectively. Note that, for synchronized transmitter and receiver, we can substitute TOA $\hat{\tau}_j$ with the distance $\hat{d}_j$ by considering the speed of light according to $\hat{d}_j=c\cdot \hat{\tau}_j$. The observations related to the $j^{\text{th}}$ NLOS path are given by 
\begin{subequations}
\begin{alignat}{3}
&\hat{d}_j=d_j+e_{d_j}=\left\|\vec{q}-\vec{s}_j\right\|+\left\|\vec{s}_j-\vec{p}\right\|+e_{d_j},
\label{eq:NLOS_TOA}
\\
&\hat{\theta}_{\mathrm{TX},j}= \theta_{\mathrm{TX},j}+e_{\theta_{\mathrm{TX},j}}=\atantwo\left(\frac{s_{\mathrm{y},j}-q_{\mathrm{y}}}{s_{\mathrm{x},j}-q_{\mathrm{x}}}\right)+e_{\theta_{\mathrm{TX},j}},
\label{eq:NLOS_AOD}
\\
&\hat{\theta}_{\mathrm{RX},j}=\theta_{\mathrm{RX},j} +e_{\theta_{\mathrm{RX},j}}=\atantwo\left(\frac{s_{\mathrm{y},j}-p_{\mathrm{y}}}{s_{\mathrm{x},j}-p_{\mathrm{x}}}\right)-\alpha+e_{\theta_{\mathrm{RX},j}},
\label{eq:NLOS_AOA}
\end{alignat}
\end{subequations}
where $\atantwo$ is the four-quadrant inverse tangent and $e_{d_j}$, $e_{\theta_{\mathrm{TX},j}}$, and $e_{\theta_{\mathrm{RX},j}}$ denote estimation errors regarding the distance, AOD, and AOA, respectively. We assume that the observations are conditionally independent \cite{SZASW2017} and the measurement noise $e_{d_j}$, $e_{\theta_{\mathrm{TX},j}}$, $e_{\theta_{\mathrm{RX},j}},\forall j$ can be modeled as Gaussian distributed with zero mean and known variances $\sigma^2_{d_j}$, $\sigma^2_{\theta_{\mathrm{TX},j}}$, and $\sigma^2_{\theta_{\mathrm{RX},j}}$ \cite{TVDW2017}, respectively. This assumption was originally introduced in \cite{TVDW2017}, where it was observed that the observation errors which resulted from the considered TOA, AOD, and AOA-estimator follow a Gaussian distribution. The variances of the observation errors generally depend on the signal-to-noise-power-ratio (SNR), the bandwidth, the antenna arrays, as well as the actual estimation algorithm. For a given estimator, these values can be obtained via simulation and stored in tables for different SNRs. Finally, we assume that the position and orientation of the mobile terminal and the points of incidence are independent of each other. 

The goal of the mobile terminal is to estimate its own position and orientation, as well as the points of incidence based on the observations $\hat{\vec{z}}$ in \eqref{eq:observation}. We summarize these parameters in the vector 
\begin{equation}
\boldsymbol\eta = [\vec{p}^{\mathrm{T}},\alpha,\vec{s}_0^{\mathrm{T}},...,\vec{s}_{J-1}^{\mathrm{T}}]^{\mathrm{T}}.
\label{eq:eta}
\end{equation}
\section{Message Passing for Joint Positioning, Orientation Estimation, and Mapping}
\label{sec:BP}
This section contains the theory required for the proposed message passing-based estimator. First, we derive the factorized a posteriori distribution (short: \textit{posterior}) and the corresponding factor graph. Based on this factor graph, we briefly review the concept of belief propagation and discuss the initialization of the message passing algorithm.
\subsection{Factorized A Posteriori Distribution}
The joint a posteriori distribution is proportional to
\begin{equation}
 p(\boldsymbol\eta | \hat{\vec{z}}) \propto p(\hat{\vec{z}}|\boldsymbol\eta ) p(\boldsymbol\eta),
\label{eq:joint_posterior}
\end{equation}
where the joint likelihood function $p(\hat{\vec{z}}|\boldsymbol\eta)$ can be factorized based on the conditional independencies described in section \ref{sec:sys_model}, i.e.
\begin{equation}
\begin{split}
 p(\hat{\vec{z}}|\boldsymbol\eta)=\prod_{j=0}^{J-1} &p(\hat{d}_j|\vec{p},\vec{s}_j,\vec{q})p(\hat{\theta}_{\mathrm{TX},j}|\vec{s}_j,\vec{q}) \\
& \times p(\hat{\theta}_{\mathrm{RX},j}|\vec{s}_j,\vec{p},\alpha),
\label{eq:joint_likelihood}
\end{split}
\end{equation}
and the joint a priori distributions (short: \textit{priors}) can be factorized as follows:
\begin{equation}
p(\boldsymbol\eta) = p(\vec{p})p(\alpha)\prod_{j=0}^{J-1}p(\vec{s}_j).
\label{eq:joint_prior}
\end{equation}
The factors related to the distance, AOD, and AOA of the $j^{\text{th}}$ NLOS path in \eqref{eq:joint_likelihood} are given by
\begin{subequations}
\begin{alignat}{3}
&p(\hat{d}_j|\vec{p},\vec{q},\vec{s}_j)\propto e^{-\left(\hat{d}_j-\left\|\vec{q}-\vec{s}_j\right\|-\left\|\vec{s}_j-\vec{p}\right\|\right)^2/2\sigma_{d_j}^2},
\label{eq:p_NLOS_TOA} 
\\
&p(\hat{\theta}_{\mathrm{TX},j}|\vec{s}_j,\vec{q})\propto e^{-\left(\hat{\theta}_{\mathrm{TX},j}-\atantwo\left(\frac{s_{\mathrm{y},j}-q_{\mathrm{y}}}{s_{\mathrm{x},j}-q_{\mathrm{x}}}\right)\right)^2/2\sigma_{\theta_{\mathrm{TX},j}}^2},
\label{eq:p_NLOS_AOD} 
\\
&p(\hat{\theta}_{\mathrm{RX},j}|\vec{s}_j,\vec{p},\alpha)\propto e^{-\left(\hat{\theta}_{\mathrm{RX},j}-\atantwo\left(\frac{s_{\mathrm{y},j}-p_{\mathrm{y}}}{s_{\mathrm{x},j}-p_{\mathrm{x}}}\right)+\alpha\right)^2/2\sigma_{\theta_{\mathrm{RX},j}}^2},
\label{eq:p_NLOS_AOA} 
\end{alignat}
\end{subequations}
respectively. The posterior distribution in \eqref{eq:joint_posterior} is non-convex and has many local maxima. Hence it is difficult to obtain optimum estimates (e.g., maximum a posteriori (MAP) estimates) since numerical solvers will get stuck in local extrema if the initial estimate is far away from the global optimum.
\begin{figure}[t]%
\centering
\includegraphics[width=.8\columnwidth]{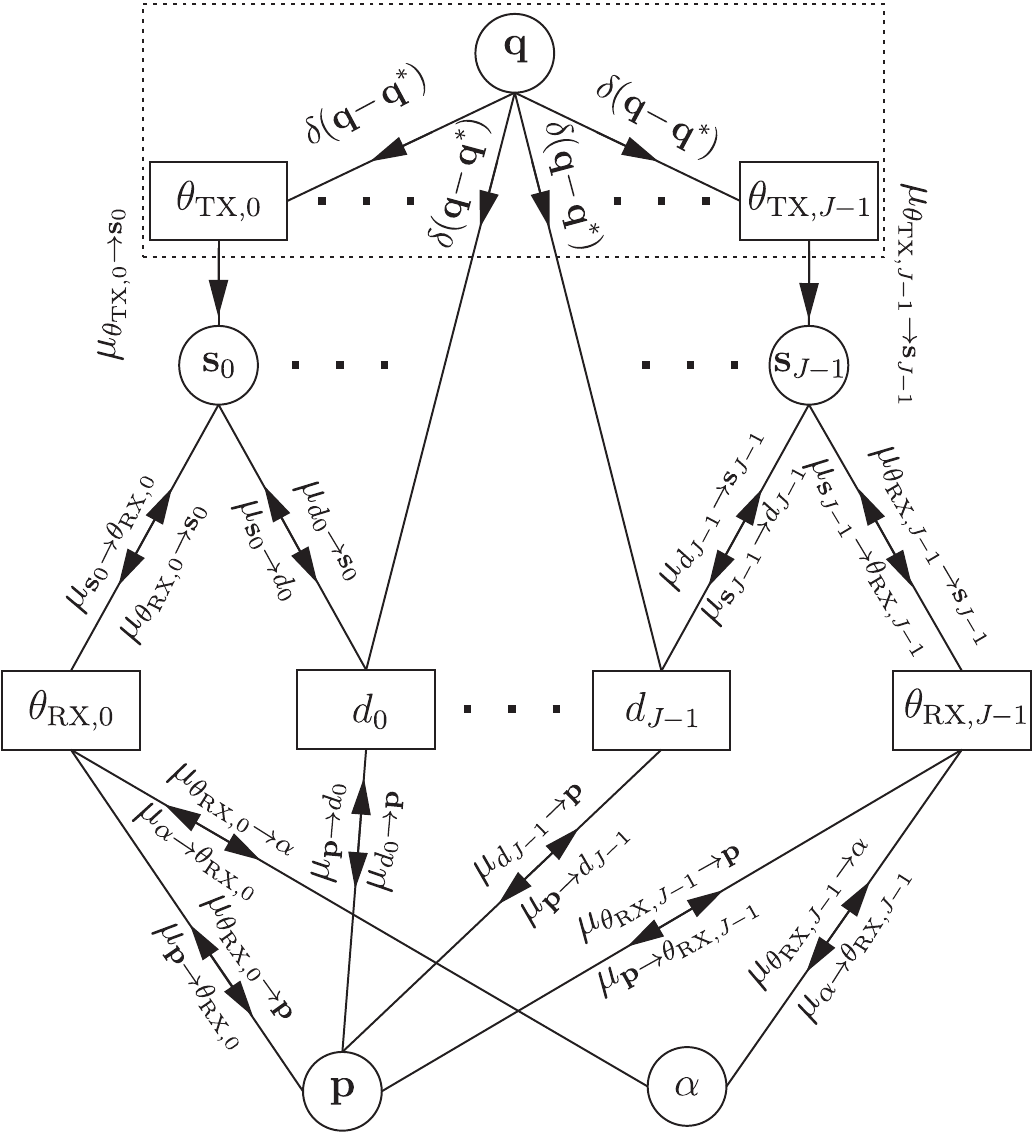}%
\label{subfig:FG_NLOS}%
\caption{\textit{Factor graph of the posterior distribution in \eqref{eq:joint_posterior}-} Messages are passed along the edges of the factor graph to iteratively determine the marginals of $\vec{p}$, $\alpha$, and $\vec{s}_j,\forall j$.}
\label{fig:FG}
\end{figure}
\subsection{Factor Graph}
We can visualize the factorized a posteriori distribution in a graphical way using the notion of factor graphs. The factor graph corresponding to the a posteriori distribution in \eqref{eq:joint_posterior} is depicted in Fig. \ref{fig:FG}. Factor graphs are bipartite graphs that consist of factor nodes (rectangles in Fig. \ref{fig:FG}), variable nodes (circles in Fig. \ref{fig:FG}), and edges to connect the nodes \cite{W2007}. 

The factor graph in Fig. \ref{fig:FG} helps us to reveal the structure of the estimation problem. For instance, we observe that any factor node $d_j$ is connected to $\vec{p}$, $\vec{q}$, and $\vec{s}_j$ meaning that the distance estimate is useless for $\vec{p}$ if we have no information regarding $\vec{s}_j$ and vice versa. Similarly, $\theta_{\mathrm{RX},j}$ becomes only useful for $\alpha$ if we have knowledge about $\vec{p}$ and $\vec{s}_j$. From the factor graph, we can deduce that we have to initialize the message passing algorithm from $\vec{q}$ via $\vec{s}_j,\forall j$ to $\vec{p}$ and $\alpha$. In other words, the information from the base station initially trickles down to the position and orientation via the points of incidence. We will use this insight in section \ref{subsec:init}, to derive an initialization strategy.
\subsection{Belief Propagation}
In contrast to numerical solvers which try to find optimum estimates based on the high-dimensional joint posterior distribution (here $\mathrm{dim}(\boldsymbol\eta)= 3+2J$), belief propagation first determines the lower-dimensional marginal posterior distributions (e.g., p($\vec{p}|\hat{\vec{z}})$ with $\mathrm{dim}(\vec{p}) = 2 \ll \mathrm{dim} (\boldsymbol\eta)$). Estimates are obtained based on the marginals subsequently. The marginals are determined iteratively by passing messages along the edges of the underlying factor graph \cite{W2007}. At all nodes of the graph, outgoing messages are updated based on the incoming messages and the type of node. Belief propagation has two main update operations, namely, \textit{message filtering} (messages from factor to variable nodes) and \textit{message multiplication} (messages from variable to factor nodes). 
\subsubsection{Message Filtering}
Every factor node computes an outgoing message for every edge based on the function related to the factor node and all incoming messages, excluding the message from the edge for which the outgoing message is computed. For instance, the message from factor node $d_j$ in Fig. \ref{fig:FG} to the variable node $\vec{s}_j$ is computed as follows \cite{W2007}
\begin{equation}
	\mu_{d_j \rightarrow \vec{s}_j}^{(l)}(\vec{s}_j)	\propto \int p(\hat{d}_j|\vec{p},\vec{q},\vec{s}_j) \times\mu_{\vec{p} \rightarrow d_j}^{(l-1)}(\vec{p})\delta(\vec{q}-\vec{q}^*)\mathrm{d} \vec{p}\mathrm{d} \vec{q},
	\label{eq:message_filtering}
\end{equation}
where the superscript $(l)$ refers to the iteration index, $p(\hat{d}_j|\vec{p},\vec{q},\vec{s}_j)$ is defined in \eqref{eq:NLOS_TOA}, $\mu_{\vec{p} \rightarrow d_j}^{(l-1)}(\vec{s}_j)$ and $\delta(\vec{q}-\vec{q}^*)$ are the incoming message from $\vec{p}$ and $\vec{q}$, respectively. Note that the integral in \eqref{eq:message_filtering} cannot be solved in closed-form unless the position $\vec{p}$ of the agent is perfectly determined, i.e. $\mu_{\vec{p} \rightarrow \theta_{d_j}}^{(l)}(\vec{s}_j)=\delta(\vec{p}-\vec{p}^{*})$. For incoming messages of generic structure, we have to resort to particle-based approximations, as will be explained in section \ref{subsec:particle_based_computation}.
\subsubsection{Message Multiplication}
Every variable node computes an outgoing message for every edge based on the product of the previous belief and all incoming messages excluding the message from the edge for which the outgoing message is computed. For instance, the message from the variable node $\vec{s}_j$ to the factor node $\theta_{\mathrm{RX},j}$ is given by
\begin{equation}
	\mu_{\vec{s}_j\rightarrow \theta_{\mathrm{RX},j}}^{(l)}(\vec{s}_j)= b_{\vec{s}_j}^{(l-1)}(\vec{s}_j)\mu_{\theta_{\mathrm{TX},j} \rightarrow \vec{s}_j}^{(l)}(\vec{s}_j)\mu_{d_j \rightarrow \vec{s}_j}^{(l)}(\vec{s}_j),
\label{eq:mesasage_to_factor_node}
\end{equation}
where $b_{\vec{s}_j}^{(l-1)}(\vec{s}_j)$ is the belief on $\vec{s}_j$ from the previous iteration, while $\mu_{\theta_{\mathrm{TX},j} \rightarrow \vec{s}_j}^{(l)}(\vec{s}_j)$ and $\mu_{d_j \rightarrow \vec{s}_j}^{(l)}(\vec{s}_j)$ are the incoming messages as depicted in Fig. \ref{fig:FG}. 

The belief of the current iteration is computed as the product of all incoming messages and the previous belief. For the previous example,
\begin{equation}
\begin{split}
b_{\vec{s}_j}^{(l)}(\vec{s}_j) 	=& b_{\vec{s}_j}^{(l-1)}(\vec{s}_j)\mu_{\theta_{\mathrm{TX},j} \rightarrow \vec{s}_j}^{(l)}(\vec{s}_j)
\\
& \times \mu_{\theta_{\mathrm{RX},j} \rightarrow \vec{s}_j}^{(l)}(\vec{s}_j)	\mu_{d_j \rightarrow \vec{s}_j}^{(l)}(\vec{s}_j).
\label{eq:belief}
\end{split}
\end{equation}

\subsection{Initialization}
\label{subsec:init}

Message passing algorithms are generally initialized by the leaf nodes of the graph \cite{W2007}. Recall that we are considering the most general case, where we have no prior information regarding $\vec{p}$, $\alpha$, and $\vec{s}_j,\forall j$. Consequently, the base station node $\vec{q}$ is the only node with a non-uniform prior and belief propagation is initialized at this node. Recall that the base station's position and orientation are perfectly known and, thus, a Dirac function $\delta(\vec{q}-\vec{q}^*)$ is passed towards all connected nodes, where $\vec{q}^*$ is the true location of the base station. Note that in the upper part of the factor graph (dashed box in Fig. \ref{fig:FG}), message-flow is unidirectional, i.e. no messages are sent back to the base station since its position is perfectly determined. We choose the following sequence of messages for initialization: 1) $\mu_{\theta_{\mathrm{TX},j}\rightarrow \vec{s}_j}, \forall j$, 2) $\mu_{ \vec{s}_j\rightarrow d_j} =\mu_{\theta_{\mathrm{TX},j}\rightarrow \vec{s}_j},\forall j$, 3) $\mu_{ d_j\rightarrow \vec{p}}, \forall j$, 4) $\mu_{ \vec{p}\rightarrow d_j}, \forall j$ and $\mu_{ \vec{p}\rightarrow \theta_{\mathrm{RX},j}},\forall j$, 5) $\mu_{ d_j\rightarrow \vec{s}_j}, \forall j$, 6) $\mu_{ \vec{s}_j\rightarrow \theta_{\mathrm{RX},j}}, \forall j$, 7)$\mu_{ \theta_{\mathrm{RX},j} \rightarrow \alpha }, \forall j$, 8) $\mu_{ \alpha \rightarrow \theta_{\mathrm{RX},j} }, \forall j$, 9) $\mu_{ \theta_{\mathrm{RX},j} \rightarrow \vec{p}}, \forall j$ and $\mu_{ \theta_{\mathrm{RX},j} \rightarrow \vec{s}_j}, \forall j$.
After this sequences of messages, the factor nodes have incoming messages from all edges, and the beliefs are determined based on these messages. In all subsequent iterations, a so-called flooding schedule is used to update messages\cite{SLG2004}. 

\textit{Remark:} Note that the initialization strategy is not unique. However, empirical observations showed that it leads to fast convergence and high accuracies of the estimates.
\section{Particle-based Message Computation and Estimation}
\label{sec:NBP}
First, we briefly review the concept of importance sampling to approximate the continuous messages by sets of weighted samples (particles). Afterwards, we explain how the continuous messages in \eqref{eq:message_filtering} and \eqref{eq:mesasage_to_factor_node} are approximated by sets of particles. Finally, we explain how to obtain estimates of the parameters based on their beliefs. For notational convenience, we drop the iteration-superscript in this section.
\subsection{Importance Sampling}
To perform belief propagation, we need means to compute the outgoing messages. The filtering operation in \eqref{eq:message_filtering} requires solving an integral which cannot be solved in closed-form in general. Since all messages can be interpreted as probability distributions, our goal is to draw samples from these distributions without computing these distributions explicitly. To that end, we employ importance sampling. 

In importance sampling, we wish to obtain a set of samples $\vec{x}^{(k)},  k=1,...,N_s$ from $p(\vec{x})$, where $p(\vec{x})$ cannot be sampled directly. In our context $p(\vec{x})$ is any outgoing message, e.g., $\mu_{\theta_{\mathrm{RX},j} \rightarrow \vec{s}_j}^{(l)}(\vec{s}_j)$. Therefore, we draw $N_s$ samples $\vec{x}^{(k)}$ from a suitable \textit{proposal distribution} $q(\mathbf{x})$ and attach a weight $w^{(k)}$ to every sample. The weight accounts for the mismatch between $p(\vec{x})$ and $q(\vec{x})$ \cite{KTB2011}. The combination of a sample and its weight is referred to  as a particle $\{w^{(k)},\vec{x}^{(k)}\}$. The unnormalized weight is given by \cite{KTB2011}
\begin{equation}
\tilde{w}^{(k)} = \frac{p(\vec{x}^{(k)})}{q(\vec{x}^{(k)})}.
\label{eq:weights}
\end{equation}
For numerical stability, we normalize all weights such that $w^{(k)}=\tilde{w}^{(k)}/\sum_k \tilde{w}^{(k)}$. The set of samples with their associated weights is called \textit{particle representation} of $p(\vec{x})$, denoted by $\mathcal{R}_{N_s}\left(p(\vec{x})\right)$. Finally, we resample the particle representation to stochastically discard particles with very low weights and reproduce particles with high weights \cite{KTB2011}.

\subsection{Particle-based Message Computation}
\label{subsec:particle_based_computation}
\subsubsection{Message Filtering}
At any factor node, assume that all incoming messages are given as particle representations and we wish to obtain a particle representation of an outgoing message. For instance, we want to obtain the particle representation $\mathcal{R}_{N_s}\left(\mu_{d_j \rightarrow \vec{s}_j}(\vec{s}_j)\right)=\{w_{\vec{s}_j}^{(k)},\vec{s}_j^{(k)}\}_{k=1}^{N_s}$ of the filtered message in \eqref{eq:message_filtering} based on the particle representation of the incoming message $\mathcal{R}_{N_s}\left(\mu_{\vec{p} \rightarrow d_j}(\vec{s}_j)\right)=\{w_{\vec{p}}^{(n)},\vec{p}^{(n)}\}_{n=1}^{N_s}$. Given a set of samples $\{\vec{s}_j^{(k)}\}_{k=1}^{N_s}$ from the proposal distribution $q(\vec{s}_j)$, the unnormalized weights are computed according to
\begin{equation}
\tilde{w}_{\vec{s}_j}^{(k)} = \frac{\mu_{d_j \rightarrow \vec{s}_j}(\vec{s}_j^{(k)})}{q(\vec{s}_j^{(k)})} = \frac{\sum_{n=1}^{N_s} w_{\vec{p}}^{(n)}p(\hat{d}_j|\vec{p}^{(n)},\vec{q}^*,\vec{s}_j^{(k)})}{q(\vec{s}_j^{(k)})},
\label{eq:weights_filtering}
\end{equation}
where the integration in \eqref{eq:message_filtering} is replaced by the sum over all particles of the incoming message $\mathcal{R}_{N_s}\left(\mu_{\vec{p} \rightarrow d_j}(\vec{s}_j)\right)=\{w_{\vec{p}}^{(n)},\vec{p}^{(n)}\}_{n=1}^{N_s}$.

\subsubsection{Message Multiplication}
At any variable node, assume that all incoming messages are given as particle representations and we wish to obtain a particle representation of an outgoing message. For instance, we want to obtain the particle representation $\mathcal{R}_{N_s}\left(\mu_{\vec{s}_j\rightarrow \theta_{\mathrm{RX},j}}(\vec{s}_j)\right)=\{w_{\vec{s}_j}^{(k)},\vec{s}_j^{(k)}\}_{k=1}^{N_s}$  of the product in \eqref{eq:mesasage_to_factor_node} based on particle representations of the incoming messages and the previous belief. Since the samples of the incoming messages and the previously belief are drawn randomly and from independent proposal distributions, they will be distinct with probability one. Direct message multiplication is therefore not possible. 

To enable multiplication, interpolated versions of these messages (so-called kernel density estimates) are determined \cite{KTB2011}. In kernel density estimation, each particle is coated with a continuous kernel and the superposition of all $N_s$ kernels yields the resulting density. For a set of particles $\{w^{(k)},\vec{x}^{(k)}\}$ from the distribution $p(\vec{x})$, a kernel density estimate $\hat{p}(\vec{x})$ is given by
\begin{equation}
\hat{p}(\vec{x})=\sum_{k=1}^{N_s}w^{(k)} \mathcal{N}(\vec{x};\vec{x}^{(k)},\sigma_{\mathrm{KDE}}^2 \vec{I}),
\label{eq:KDE_message}
\end{equation}
where $\mathcal{N}(\vec{x};\vec{x}^{(k)},\sigma_{\mathrm{KDE}}^2 \vec{I})$ denotes the Gaussian distribution with mean $\vec{x}^{(k)}$ and covariance matrix $\sigma_{\mathrm{KDE}}^2 \vec{I}$.

Using kernel density estimates of the incoming messages and the previous belief, the current belief can also be determined using importance sampling. In particular, we draw a set of samples $\{\vec{s}_j^{(k)}\}_{k=1}^{N_s} \sim q(\vec{s}_j)$ and adjust the weights according to
\begin{equation}
\tilde{w}_{\vec{s}_j} = \frac{\hat{b}_{\vec{s}_j}(\vec{s}_j^{(k)}) \hat{\mu}_{\theta_{\mathrm{TX},j} \rightarrow \vec{s}_j}(\vec{s}_j^{(k)})\hat{\mu}_{\tau_j \rightarrow \vec{s}_j}(\vec{s}_j^{(k)})}{q(\vec{s}_j^{(k)})}.
\label{eq:weights_belief}
\end{equation}
Due to space limitations, we provided an extended version of this paper online to visualize the messages passed along the edges of the factor graph. Please refer to \cite{MWB2018b} for a descriptive illustration of the messages. 
\subsection{Estimation and Implementation Considerations}
In every iteration, we obtain estimates on the position and orientation of the mobile terminal, as well as the points of incidence based on their beliefs. Since the beliefs are given as particle representation, an MMSE estimate can be obtained by computing the centroid of the cloud of particles \cite{WLW2009}. For instance, assume that the belief of the $j^{\mathrm{th}}$ scatterer is given by the set particles $\mathcal{R}_{N_s}\left(b_{\vec{s}_j}(\vec{s}_j)\right)=\{w_{\vec{s}_j}^{(k)},\vec{s}_j^{(k)}\}_{k=1}^{N_s}$. The MMSE estimate $\hat{\vec{s}}_{j,\mathrm{MMSE}}$ is given by 
\begin{equation}
\hat{\vec{s}}_{j,\mathrm{MMSE}}=\sum_{k=1}^{N_s}{w_{\vec{s}_j}^{(k)}} \vec{s}_j^{(k)}.
\label{eq:MMSE_estimate}
\end{equation}

For the implementation of the algorithm, we have to carefully consider two aspects: (i) choice of the proposal distributions and (ii) choice of the kernel width $\sigma_{\mathrm{KDE}}$. Regarding (i), our goal is to draw samples in areas where a target distribution (from which we cannot sample directly) has significant probability mass. Samples which reside in regions with negligible probability mass will be assigned a weight that is close to zero. Eventually, with a high probability, these particles will be discarded after resampling. However, it is generally unknown where a target distribution has significant probability mass. Hence we use proposal distributions which draw samples uniformly inside an area of interest. For instance, the true position $\vec{p}^*$ is with high probability inside the disk with radius $r=\argmax_{j} \hat{d}_j$ centered around the base station $\vec{q}^*$. This disk is used as the area of interest of $\vec{p}$. 

Regarding the kernel width $\sigma_{\mathrm{KDE}}$, we use a set of heuristic values. In general, $\sigma_{\mathrm{KDE}}$ can be determined using kernel density estimation algorithms \cite{KTB2011}. However, we found that many of such algorithms fail to determine appropriate kernel widths which lead to convergence of the message passing algorithm. 
\section{Numerical Examples}
\label{sec:numerical_example}
To assess the performance of our estimator, we performed simulations to (i) determine the speed of convergence and investigate the impact of the number of samples and (ii) examine the effect of varying measurement noise. For that purpose, we consider a scenario with $J=3$ NLOS paths, where the points of incidence are spatially correlated in the AOD-domain: $\vec{s}_1^*=[20,10]^T$ m, $\vec{s}_2^*=[80,-10]^T$ m, and $\vec{s}_3^*=[40,0]^T$ m. The mobile terminal is located at $\vec{p}^*=[70,70]^T$ m and rotated by $\alpha=45^{\circ}$. As the performance metric, we consider the RMSE. We estimate the RMSE using 1000 Monte Carlo trials. We treat the error of the points of incidence jointly, i.e. $\vec{e}_{\vec{s}}=[\hat{\vec{s}}_{0,\mathrm{MMSE}}^T,\hdots,\hat{\vec{s}}_{J-1,\mathrm{MMSE}}^T]^T-[(\vec{s}_0^{*})^T,\hdots,(\vec{s}_{J-1}^{*})^T]^T$. As performance benchmark, we consider the LS approach from  \cite{SGDSW2017}. Recall that numerous LS solvers work in parallel each of which uses a different trial value $\alpha_{\mathrm{trial}}$. For tough comparison, we choose a very fine grid of $\Delta\alpha_{\mathrm{trial}}=0.01 \text{ rad}$. Hence we solve 629 LS problems in parallel.
\subsection{Convergence and Number of Samples}
\begin{figure}[t]%
\centering
\includegraphics[width=0.98\columnwidth]{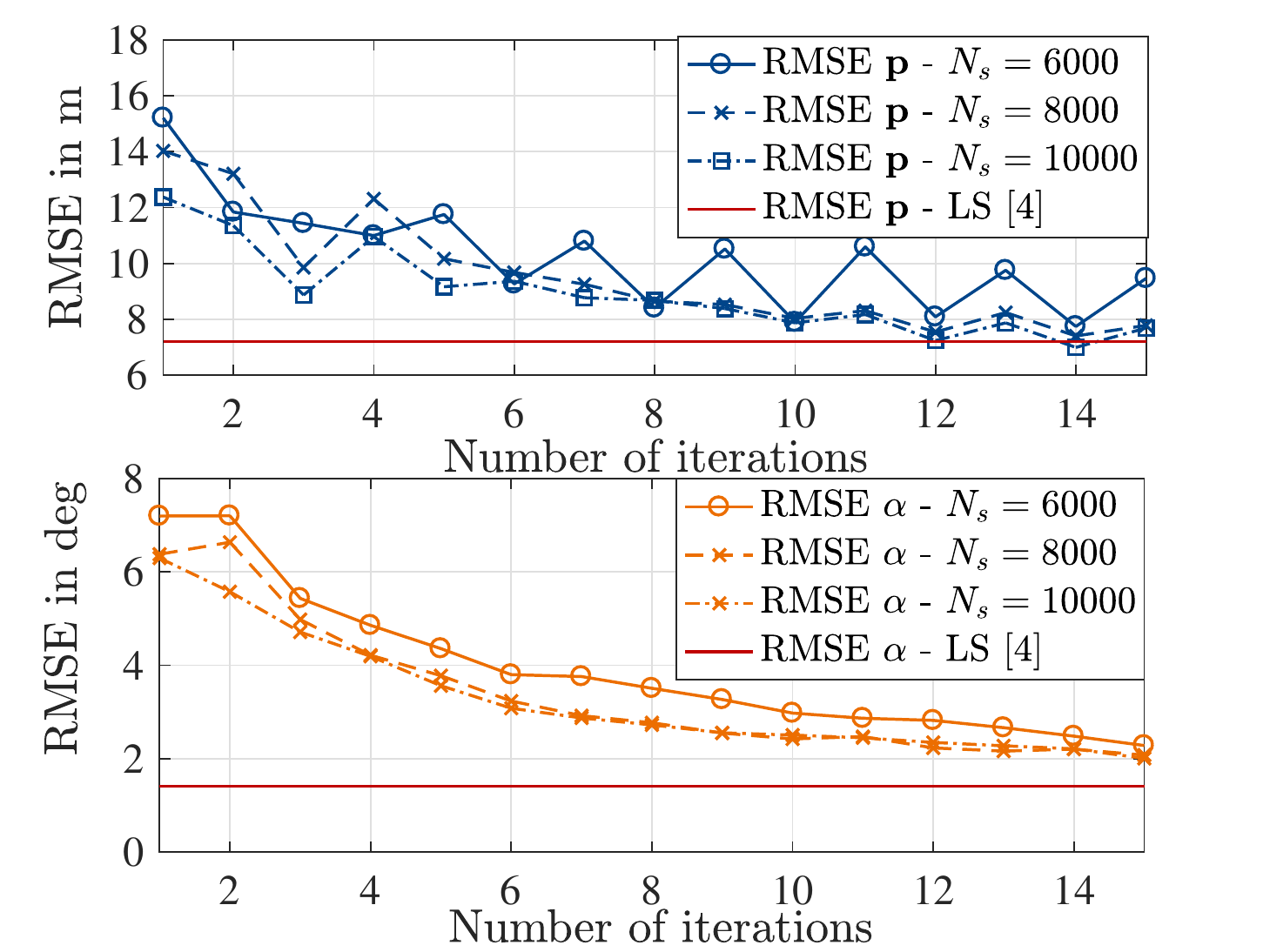}%
\caption{\textit{RMSE of $\vec{p}$ and $\alpha$ versus number of iterations} - The RMSE of position and orientation estimates decreases with the number of iterations $N_{\mathrm{Niter}}$. Increasing the number of samples results in higher estimation accuracy.}%
\label{fig:RMSE_p_alpha_Niter}%
\end{figure}

Fig. \ref{fig:RMSE_p_alpha_Niter} depicts the RMSE of the position (top) and orientation (bottom) estimates against the number of iterations for fixed measurement noise ($\sigma_{\theta_{\mathrm{TX},k}}=\sigma_{\theta_{\mathrm{RX},k}}=1^{\circ}$ and $\sigma_{d_k}=0.2$ m). For comparison, we also depict the performance of the LS estimator (solid, red horizontal lines). In both cases, the RMSE reduces with the number of iterations. The largest reduction of the RMSE occurs in the first few iterations. Note that the decrease is not monotonic. Especially, the RMSE of $\vec{p}$ shows some oscillating behavior which results from the flooding schedule mentioned in section \ref{subsec:init}. Other schedules have to be investigated in future works to mitigate the oscillation. In addition, we observe that the estimation accuracy increases with the number of samples which gives rise to a complexity-accuracy trade-off.

\subsection{Varying Measurement Noise}
\begin{figure}[t]%
\centering
\includegraphics[width=0.98\columnwidth]{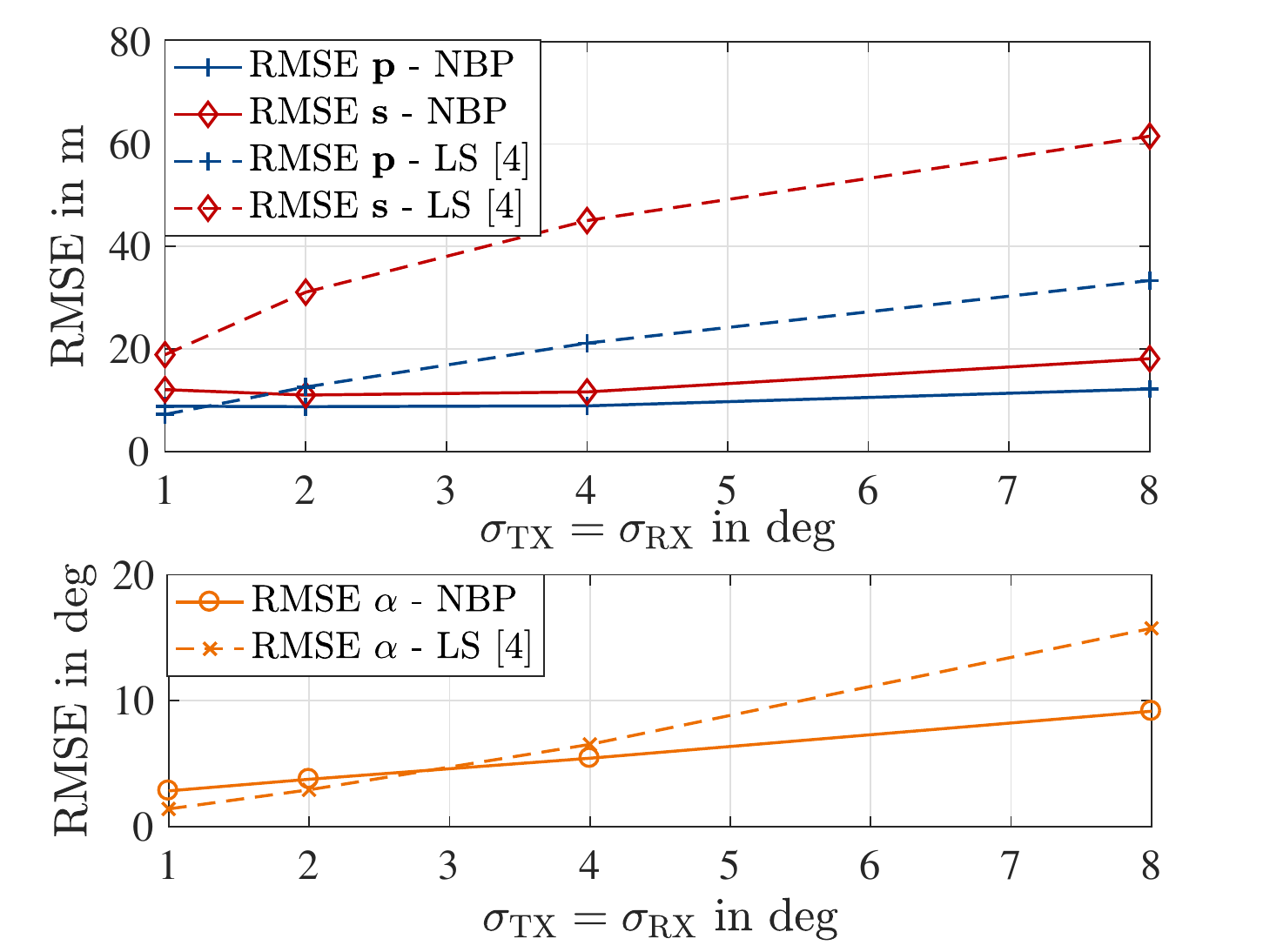}%
\caption{\textit{RMSE of $\vec{p}$, $\vec{s}$, and $\alpha$ versus angular measurement noise} - The RMSE of the LS estimates increases drastically with increasing $\sigma_{\theta_{\mathrm{TX},k}}=\sigma_{\theta_{\mathrm{RX},k}}$, while the RMSE of the NBP estimates remains almost constant.}%
\label{fig:RMSE_p_sk_alpha}%
\end{figure}

Fig. \ref{fig:RMSE_p_sk_alpha} depicts the RMSE of the position, point of incidence, and orientation estimates considering increasing angular measurement noise ($\sigma_{\theta_{\mathrm{TX},k}}=\sigma_{\theta_{\mathrm{RX},k}}\uparrow$ and fixed $\sigma_{\tau_k}=0.2$ m). We compare the RMSE of our message passing-based estimator to the RMSE of the LS approach in \cite{SGDSW2017}. We choose $N_s=10000$ and $N_{\mathrm{Niter}}=6$. Note that higher $N_{\mathrm{Niter}}$ would reduce the RMSE of our estimator further as can be seen in Fig. \ref{fig:RMSE_p_alpha_Niter}. However, the complexity would also increase. We observe in Fig. \ref{fig:RMSE_p_sk_alpha} that, for $\sigma_{\theta_{\mathrm{TX},k}}=\sigma_{\theta_{\mathrm{RX},k}}>2^{\circ}$, our proposed estimator provides significantly lower RMSE compared to the LS approach. Especially, points of incidence can be estimated more precisely. When $\sigma_{\theta_{\mathrm{TX},k}}=\sigma_{\theta_{\mathrm{RX},k}}$ is large the initial estimates of the LS solver are far from the global minimum and the solver tends to converge to a local minimum leading to its poor performance in terms of the RMSE (see Fig. \ref{fig:RMSE_p_sk_alpha}). 

\section{Conclusions}
\label{sec:conclusion}
We proposed a novel message passing-based estimator for 5G millimeter wave MIMO systems which jointly estimates the position and orientation of a mobile terminal as well as the locations of scatterers or reflectors based on distance, angle-of-departure, and angle-of-arrival measurements of all multipath components. Our estimator determines the position and orientation of a mobile terminal accurately, while simultaneously generating a precise map of the radio environment. Even in the absence of the LOS component and without assuming prior knowledge on any of the parameters, the position, orientation, and the locations of scatterers or reflectors are estimated reliably. Our approach also provides a measure of uncertainty of the estimates since it approximates the marginal a posteriori distributions of the parameters. For large measurement noise, our proposed algorithm performs very well in terms of the estimation accuracy and outperforms the state-of-the-art least squares approach. 

\appendix
\section{Visualization of the Filtered Messages}
In the following, we display the filtered messages (messages from squares to circles in Fig. \ref{fig:FG}) in the $1^{\mathrm{st}}$ and $5^{\mathrm{th}}$ iteration. To alleviate the necessity to display the weights of particles, particle representations are resampled such that all particles have the same weight. Particles are represented by colored dots.

\subsection{Messages to the Points of Incidence}
\begin{figure}%
\centering
\includegraphics[width=.45\columnwidth]{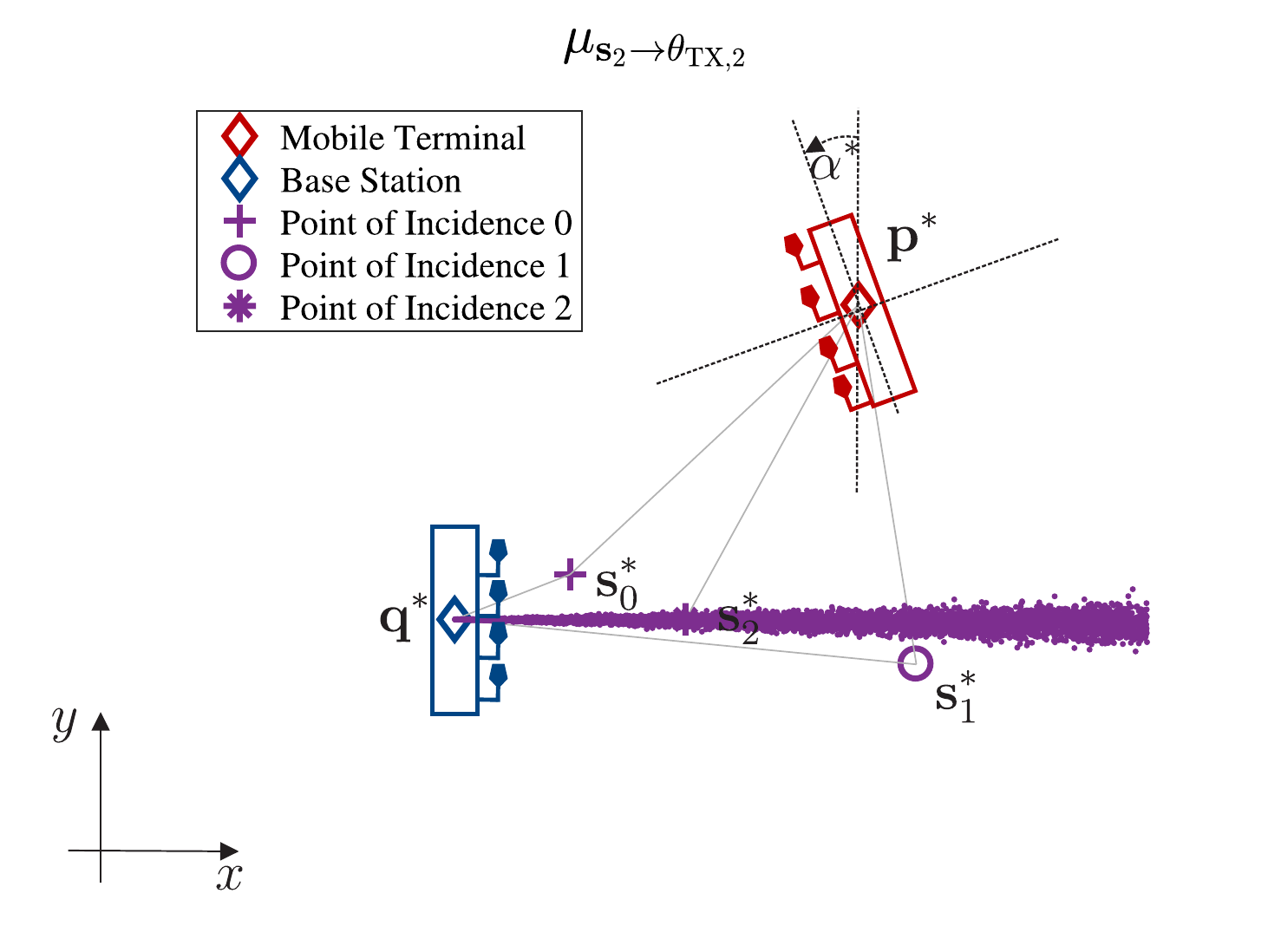}%
\caption{$\mu_{\theta_{\mathrm{TX},2} \rightarrow \vec{s}_2}$ }
\label{fig:mu_AODk_sk}
\end{figure}

In this subsection, we display the messages to nodes in the factor graph that correspond to the point of incidence $\vec{s}_2$. We start with the message $\mu_{\theta_{\mathrm{TX},2} \rightarrow \vec{s}_2}$  from AOD $\theta_{\mathrm{TX},2}$ to point of incidence $\vec{s}_2$ which is given by a cone originating at $\vec{q}$ with mean angle $\theta_{\mathrm{TX},2}$. Since the position of the base station $\vec{q}$ is perfectly known the message does not change over the iterations and the iteration superscript is omitted. This message is depicted in Fig. \ref{fig:mu_AODk_sk}.
\begin{figure*}%
\centering
\hspace*{\fill}%
\begin{subfigure}{.45\columnwidth}
\centering
\includegraphics[width=\columnwidth]{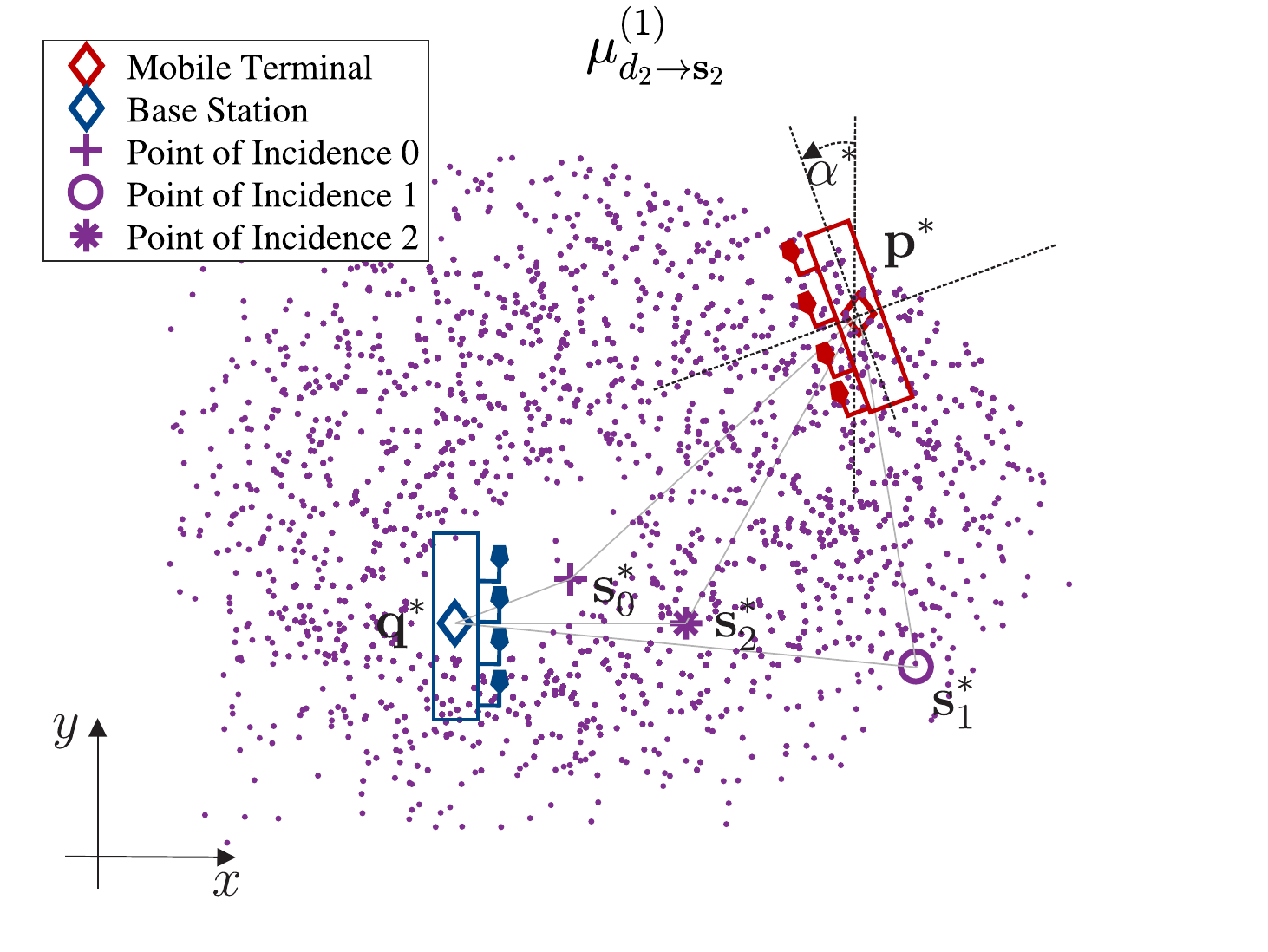}%
\label{subfig:mu_dk_sk_1st}%
\end{subfigure}
\begin{subfigure}{.45\columnwidth}
\centering
\includegraphics[width=\columnwidth]{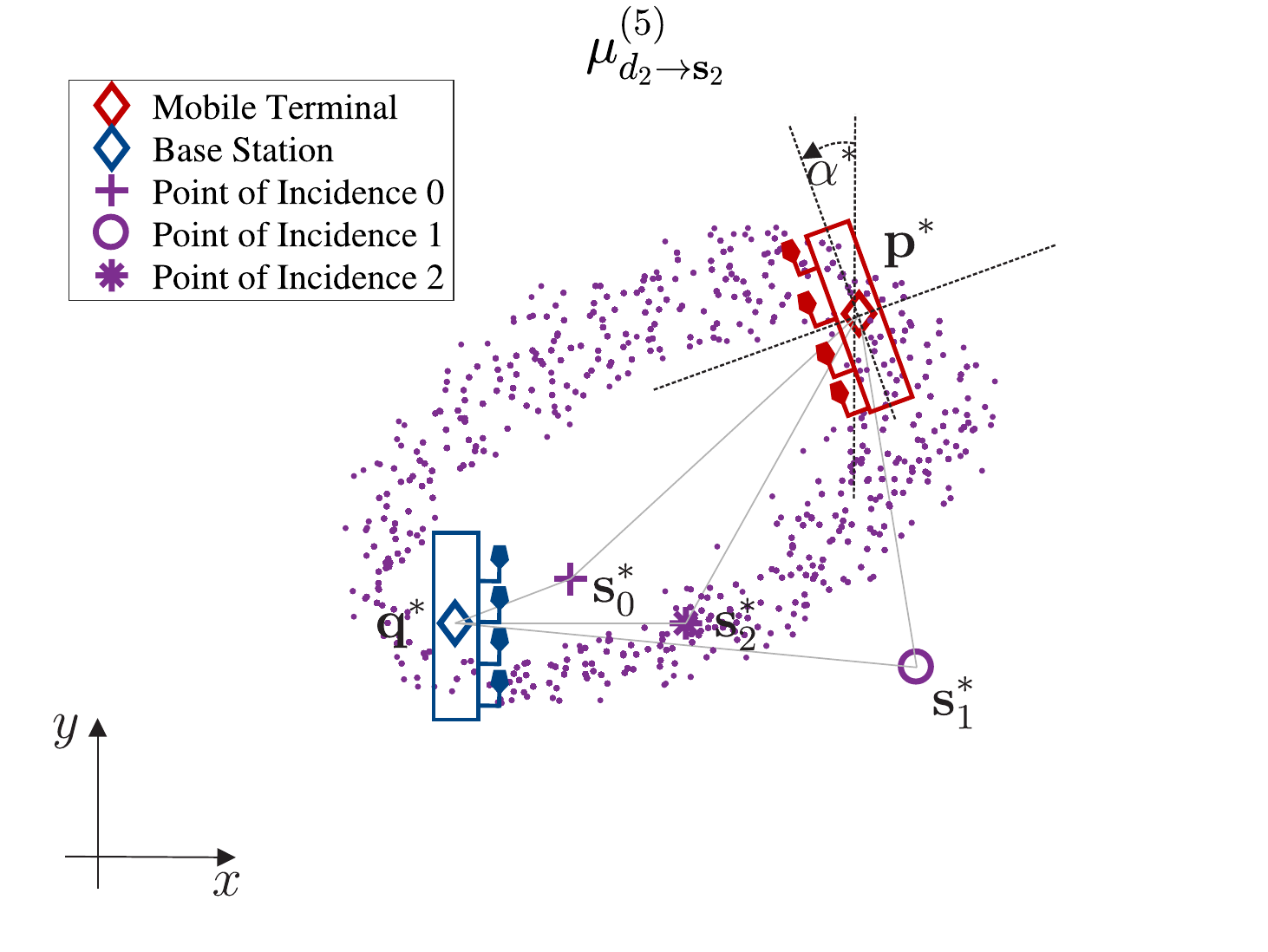}%
\label{subfig:mu_dk_sk_5th}%
\end{subfigure}
\hspace*{\fill}%
\hspace*{\fill}%
\caption{First iteration: $\mu_{d_2 \rightarrow \vec{s}_2}^{(1)}$ (left), 5th iteration: $\mu_{d_2 \rightarrow \vec{s}_2}^{(5)}$ (right) }
\label{fig:mu_dk_sk}
\end{figure*}
\begin{figure*}%
\centering
\hspace*{\fill}%
\begin{subfigure}{.45\columnwidth}
\centering
\includegraphics[width=\columnwidth]{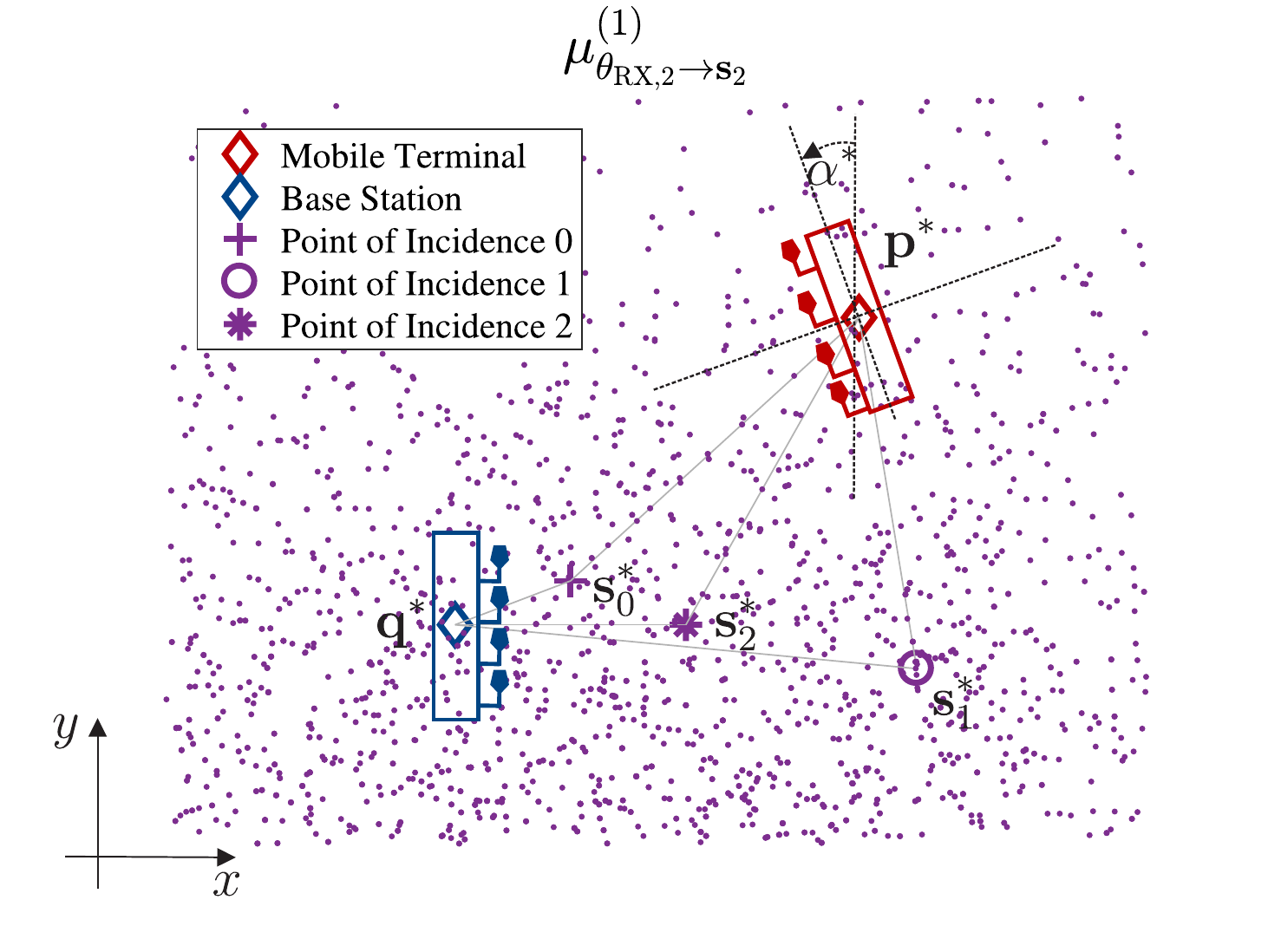}%
\label{subfig:mu_AOAk_sk_1st_it}%
\end{subfigure}
\begin{subfigure}{.45\columnwidth}
\centering
\includegraphics[width=\columnwidth]{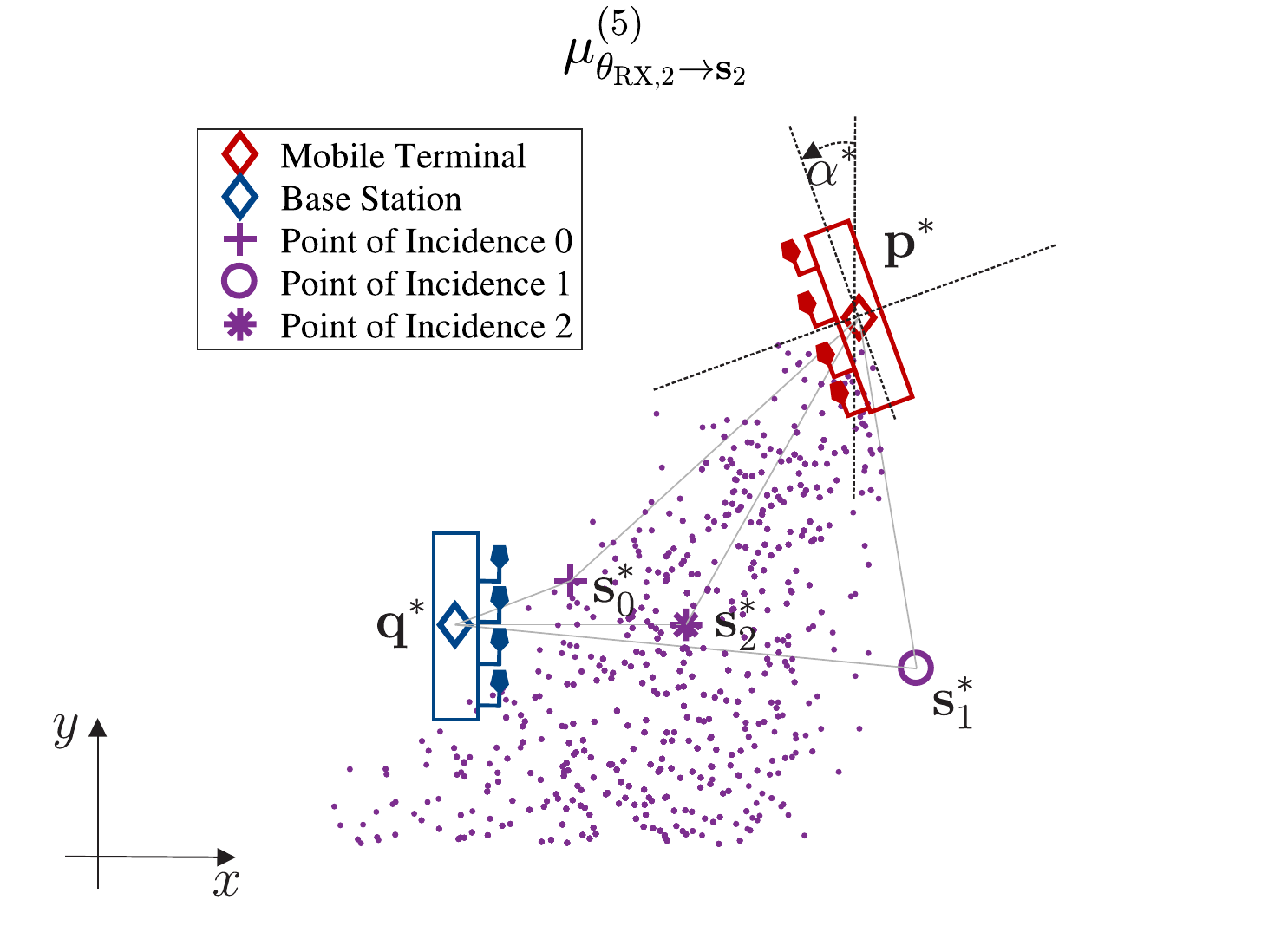}%
\label{subfig:mu_AOAk_sk_5th_it}%
\end{subfigure}
\hspace*{\fill}%
\hspace*{\fill}%
\caption{First iteration: $\mu_{\theta_{\mathrm{RX},2} \rightarrow \vec{s}_2}^{(1)}$ (left), 5th iteration: $\mu_{\theta_{\mathrm{RX},2}  \rightarrow \vec{s}_2}^{(5)}$ (right) }
\label{fig:mu_AOAk_sk}
\end{figure*}

The incoming message from the distance $d_2$ to the point of incidence $\vec{s}_2$ is depicted in Fig. \ref{fig:mu_dk_sk}. For known position $\vec{p}$, this message is given by an ellipse centered at $1/2(\vec{p}^*-\vec{q}^*)$ with a rotation of $\mathrm{atan}((p_y-q_y)/(p_x-q_x))$. When the uncertainty of $\vec{p}$ is large (e.g., in the first iteration), the 'eye' of the ellipse closes. 

The message $\mu_{\theta_{\mathrm{RX},2} \rightarrow \vec{s}_2}$  from the AOA $\theta_{\mathrm{RX},2}$ to the point of incidence  $\vec{s}_2$ is plotted in Fig. \ref{fig:mu_AOAk_sk}. When $\vec{p}$ and $\alpha$ are known the message is given by a cone originating at $\vec{p}$ with mean angle $\theta_{\mathrm{RX},2}-\alpha$.

\subsection{Messages to the Position}
In this subsection, we display a selection of the messages which arrive at the variable node $\vec{p}$. We focus on the message originating from the point of incidence $\vec{s}_2$. 

\begin{figure*}[h]%
\centering
\hspace*{\fill}%
\begin{subfigure}{.43\columnwidth}
\centering
\includegraphics[width=\columnwidth]{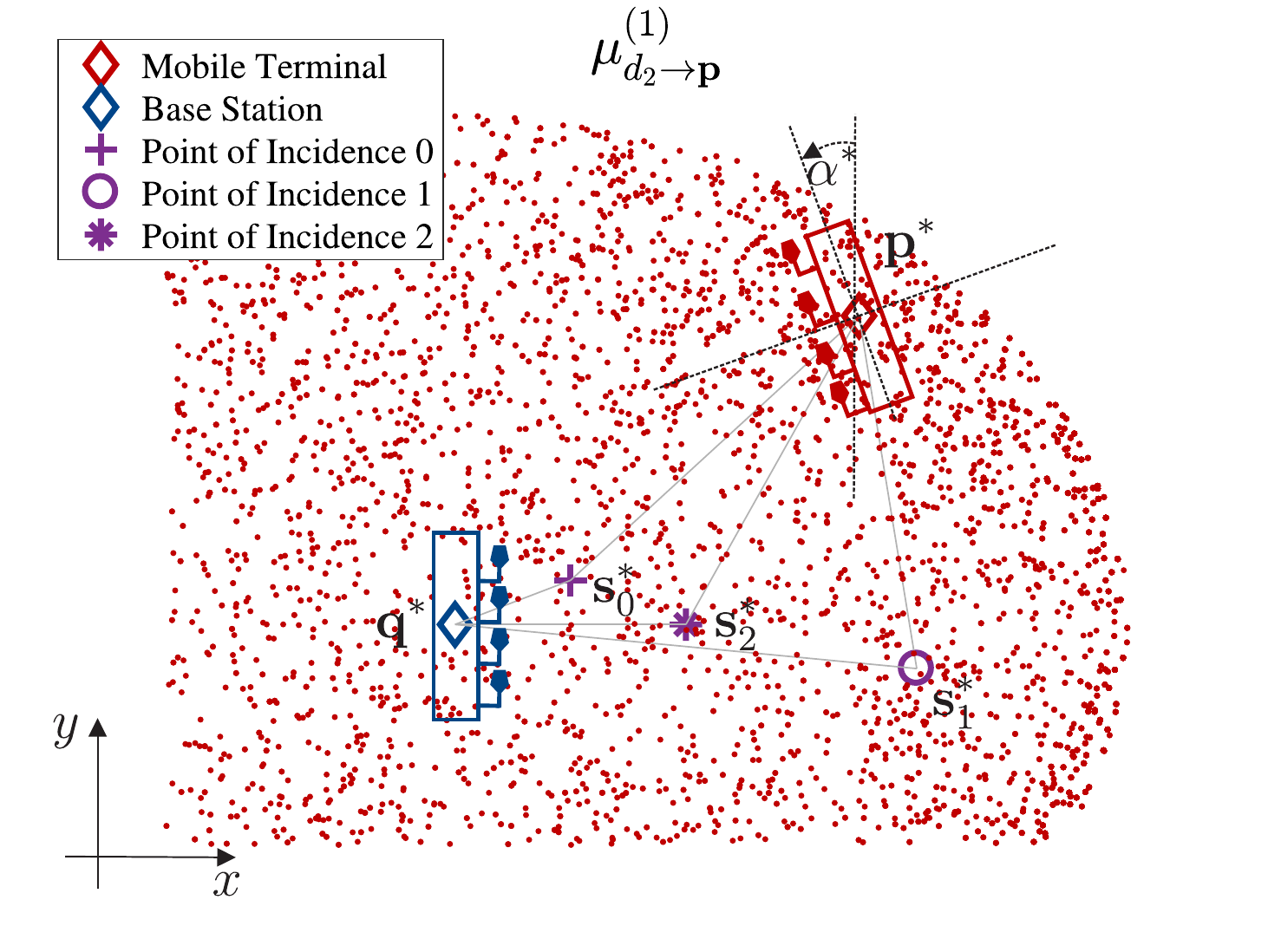}%
\label{subfig:mu_dk_p_1st_it}%
\end{subfigure}
\begin{subfigure}{.43\columnwidth}
\centering
\includegraphics[width=\columnwidth]{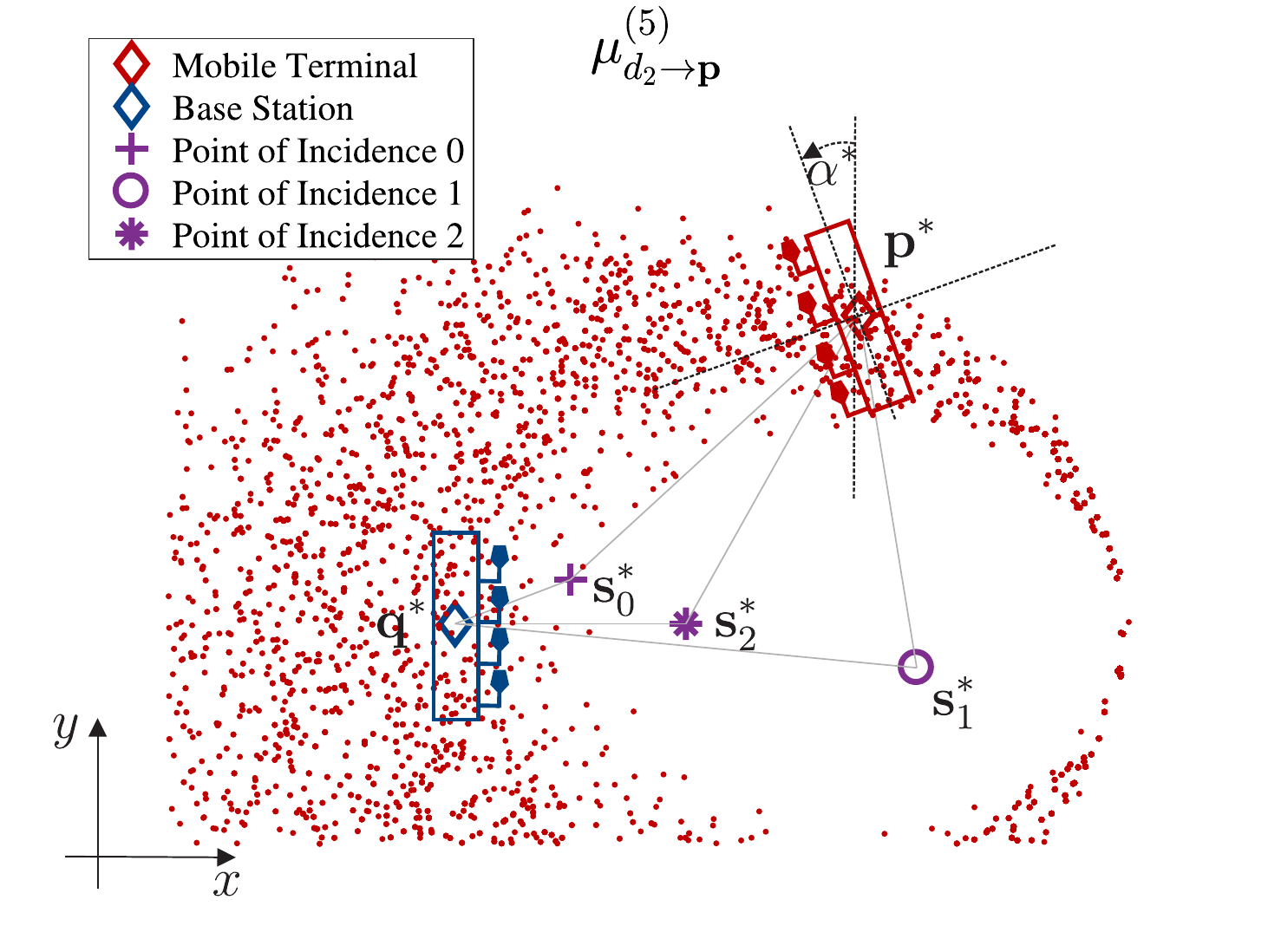}%
\label{subfig:mu_dk_p_5th_it}%
\end{subfigure}
\hspace*{\fill}%
\hspace*{\fill}%
\caption{First iteration: $\mu_{d_2 \rightarrow \vec{p}}^{(1)}$ (left), 5th iteration: $\mu_{d_2 \rightarrow \vec{p}}^{(5)}$ (right) }
\label{fig:mu_dk_p}
\end{figure*}

We start with the message from the distance $d_2$ to the position $\vec{p}$. When the location of the point of incidence $\vec{s}_2$ is perfectly known, the message is given by a circle centered at $\vec{s}_2$ with radius $d_2 - \left\|\vec{p}-\vec{q}\right\|$. Fig. \ref{fig:mu_dk_p} depicts the message $\mu_{d_2 \rightarrow \vec{p}}$ in the $1^{\mathrm{st}}$ and $5^{\mathrm{th}}$ iteration.

\begin{figure*}[h]%
\centering
\hspace*{\fill}%
\begin{subfigure}{.43\columnwidth}
\centering
\includegraphics[width=\columnwidth]{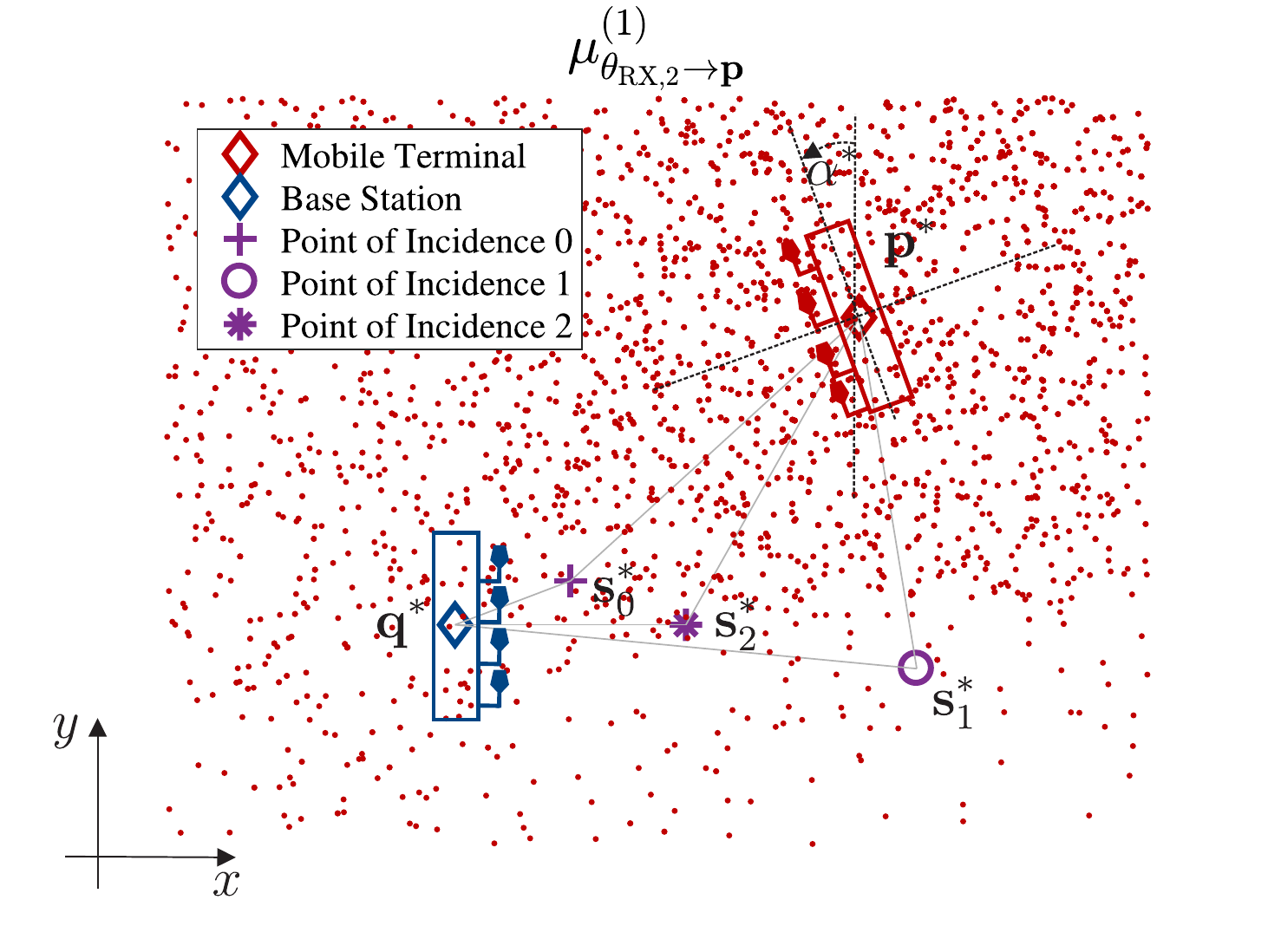}%
\label{subfig:mu_AOAk_p_1st_it}%
\end{subfigure}
\begin{subfigure}{.43\columnwidth}
\centering
\includegraphics[width=\columnwidth]{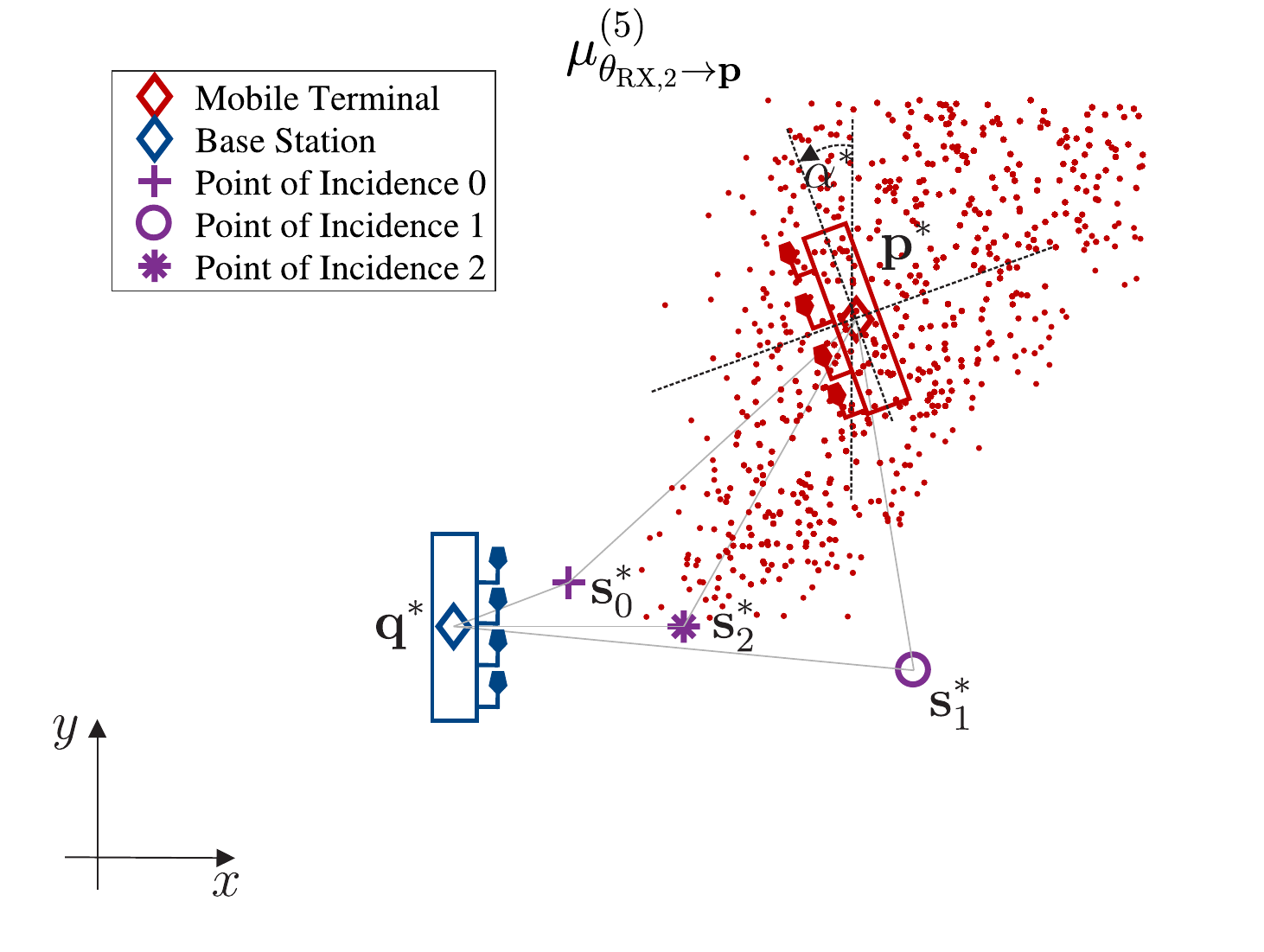}%
\label{subfig:mu_AOAk_p_5th_it}%
\end{subfigure}
\hspace*{\fill}%
\hspace*{\fill}%
\caption{First iteration: $\mu_{\theta_{\mathrm{RX},2} \rightarrow \vec{p}}^{(1)}$ (left), 5th iteration: $\mu_{\theta_{\mathrm{RX},2} \rightarrow \vec{p}}^{(5)}$ (right) }
\label{fig:mu_AOAk_p}
\end{figure*}

The message $\mu_{\theta_{\mathrm{RX},2} \rightarrow \vec{p}}$ is depicted in Fig. \ref{fig:mu_AOAk_p}. When $\vec{s}_2$ and $\alpha$ are perfectly known, the message is a cone originating at $\vec{s}_2$ with mean angle $\theta_{\mathrm{RX},2}-\alpha+\pi$. 
\subsection{Messages to the Orientation}
\begin{figure*}[h]%
\centering
\hspace*{\fill}%
\begin{subfigure}{.45\columnwidth}
\centering
\includegraphics[width=\columnwidth]{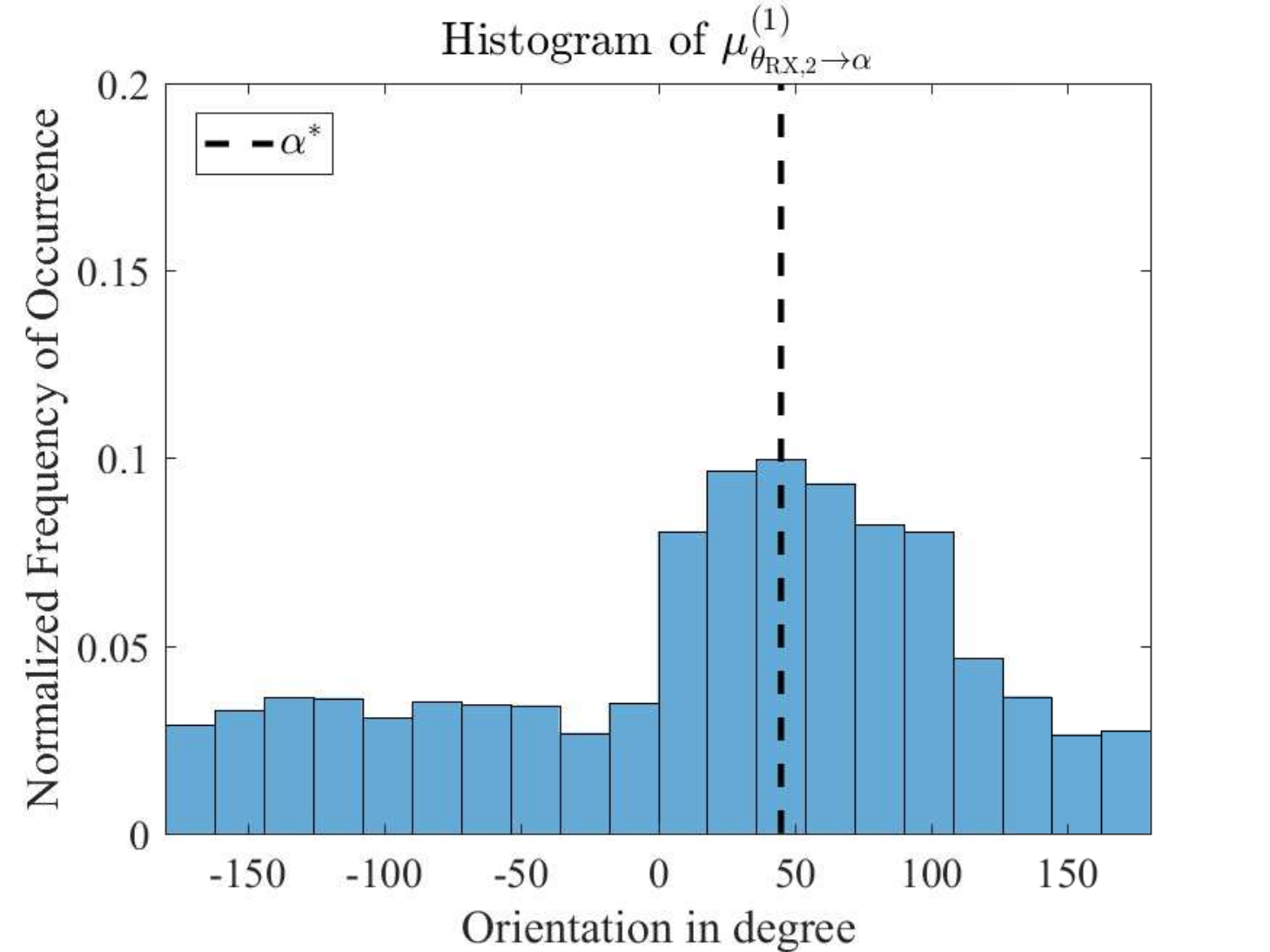}%
\label{subfig:mu_AOAk_alpha_1st_it}%
\end{subfigure}
\begin{subfigure}{.45\columnwidth}
\centering
\includegraphics[width=\columnwidth]{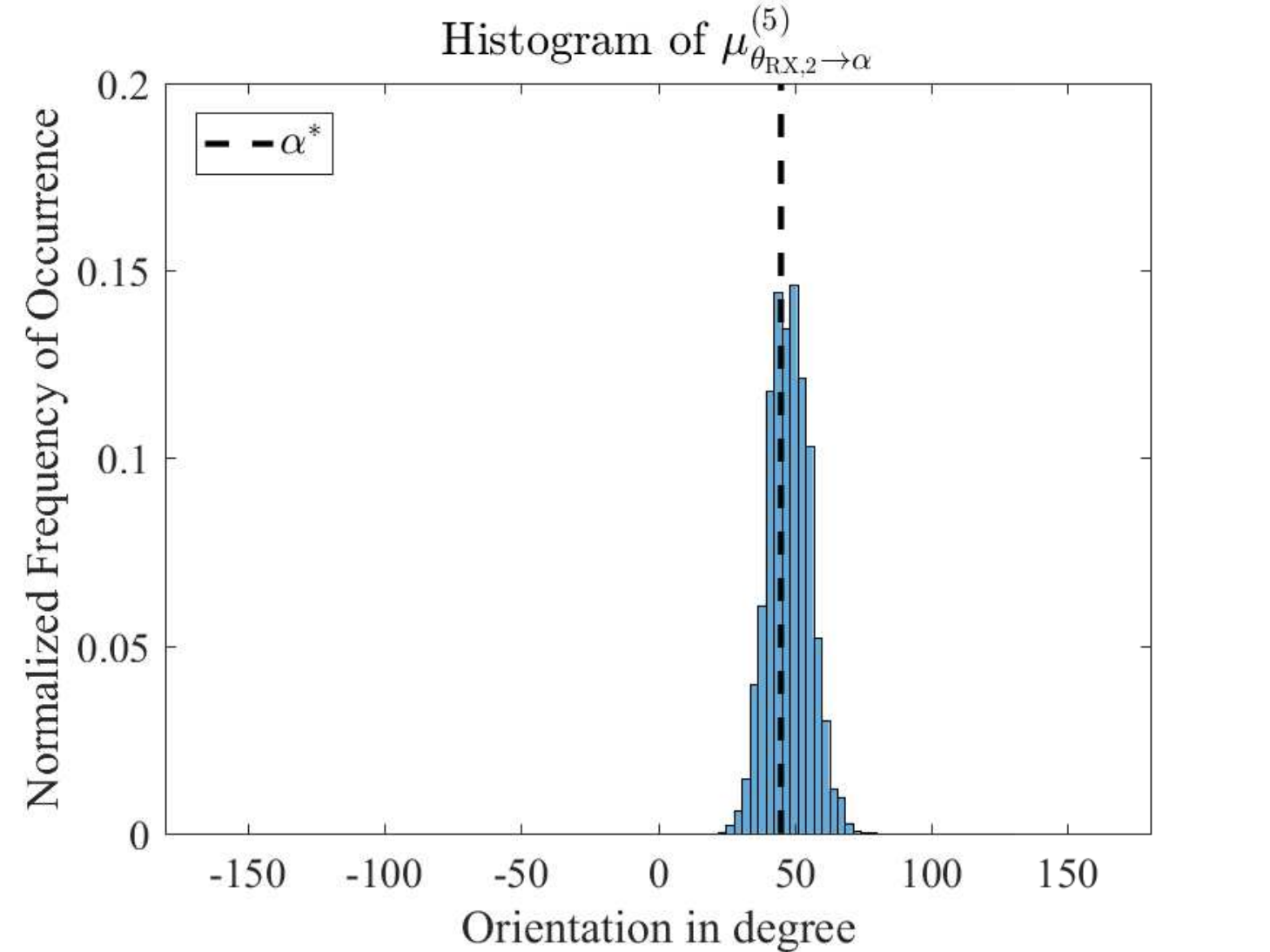}%
\label{subfig:mu_AOAk_alpha_5th_it}%
\end{subfigure}
\hspace*{\fill}%
\hspace*{\fill}%
\caption{First iteration: $\mu_{\theta_{\mathrm{RX},2} \rightarrow \alpha}^{(1)}$ (left), 5th iteration: $\mu_{\theta_{\mathrm{RX},2} \rightarrow \alpha}^{(5)}$ (right) }
\label{fig:mu_AOAk_alpha}
\end{figure*}

We depict a histogram of the message $\mu_{\theta_{\mathrm{RX},2} \rightarrow \alpha}$  from the AOA $\theta_{\mathrm{RX},2}$ to the orientation $\mu_{\theta_{\mathrm{RX},2} \rightarrow \alpha}$ in Fig. \ref{fig:mu_AOAk_alpha}. When $\vec{p}$ and $\vec{s}_2$ are perfectly known and the measurement noise of $\theta_{\mathrm{RX},2}$ obeys a zero-mean Gaussian distribution with variance $\sigma^2_{\mathrm{RX}}$, the message is given by a Gaussian distribution with mean $\alpha^*$ and variance $\sigma^2 _{\mathrm{RX}}$, where $\alpha^*$ is the true orientation. 
\IEEEpeerreviewmaketitle
\bibliographystyle{IEEEtran}
\bibliography{IEEEabrv,../../../../bibliography}

 \newcommand{\noop}[1]{}
\begin{thebibliography}{10}
\providecommand{\url}[1]{#1}
\csname url@samestyle\endcsname
\providecommand{\newblock}{\relax}
\providecommand{\bibinfo}[2]{#2}
\providecommand{\BIBentrySTDinterwordspacing}{\spaceskip=0pt\relax}
\providecommand{\BIBentryALTinterwordstretchfactor}{4}
\providecommand{\BIBentryALTinterwordspacing}{\spaceskip=\fontdimen2\font plus
\BIBentryALTinterwordstretchfactor\fontdimen3\font minus
  \fontdimen4\font\relax}
\providecommand{\BIBforeignlanguage}[2]{{%
\expandafter\ifx\csname l@#1\endcsname\relax
\typeout{** WARNING: IEEEtran.bst: No hyphenation pattern has been}%
\typeout{** loaded for the language `#1'. Using the pattern for}%
\typeout{** the default language instead.}%
\else
\language=\csname l@#1\endcsname
\fi
#2}}
\providecommand{\BIBdecl}{\relax}
\BIBdecl

\bibitem{CTXC2004}
X.~Cheng, A.~Thaeler, G.~Xue, and D.~Chen, ``{TPS: A Time-based Positioning
  Scheme for Outdoor Wireless Sensor Networks},'' in \emph{IEEE INFOCOM 2004},
  vol.~4, March 2004, pp. 2685--2696 vol.4.

\bibitem{G1996}
L.~C. Godara, ``{Limitations and Capabilities of Directions-of-Arrival
  Estimation Techniques using an Array of Antennas: A Mobile Communications
  Perspective},'' in \emph{Proceedings of International Symposium on Phased
  Array Systems and Technology}, Oct 1996, pp. 327--333.

\bibitem{SGDSW2015}
A.~Shahmansoori, G.~E. Garcia, G.~Destino, G.~Seco-Granados, and H.~Wymeersch,
  ``{5G} position and orientation estimation through millimeter wave {MIMO},''
  in \emph{2015 IEEE Globecom Workshops (GC Wkshps)}, Dec 2015, pp. 1--6.

\bibitem{SGDSW2017}
\BIBentryALTinterwordspacing
------. (2017) {Position and Orientation Estimation through Millimeter Wave
  MIMO in 5G Systems}. [Online]. Available:
  \url{{https://arxiv.org/pdf/1702.01605}}
\BIBentrySTDinterwordspacing

\bibitem{SZASW2017}
\BIBentryALTinterwordspacing
Z.~Abu-Shaban, X.~Zhou, T.~Abhayapala, G.~Seco-Granados, and H.~Wymeersch.
  (2017) {Error Bounds for Uplink and Downlink 3D Localization in 5G mmWave
  Systems}. [Online]. Available: \url{{https://arxiv.org/abs/1704.03234}}
\BIBentrySTDinterwordspacing

\bibitem{MWBA2017}
\BIBentryALTinterwordspacing
R.~Mendrzik, H.~Wymeersch, G.~Bauch, and Z.~Abu-Shaban. (2017) {Harnessing NLOS
  Components for Position and Orientation Estimation in 5G Millimeter Wave
  MIMO}. [Online]. Available: \url{{https://arxiv.org/abs/1712.01445}}
\BIBentrySTDinterwordspacing

\bibitem{WWS2017}
Y.~Wang, Y.~Wu, and Y.~Shen, ``Multipath effect mitigation by joint
  spatiotemporal separation in large-scale array localization,'' in
  \emph{GLOBECOM 2017 - 2017 IEEE Global Communications Conference}, Dec 2017,
  pp. 1--6.

\bibitem{TVDW2017}
J.~Talvitie, M.~Valkama, G.~Destino, and H.~Wymeersch, ``Novel algorithms for
  high-accuracy joint position and orientation estimation in 5g mmwave
  systems,'' in \emph{2017 IEEE Globecom Workshops (GC Wkshps)}, Dec 2017, pp.
  1--7.

\bibitem{URGW2016}
\BIBentryALTinterwordspacing
M.~Ulmschneider, R.~Raulefs, C.~Gentner, and M.~Walter, ``Multipath assisted
  positioning in vehicular applications,'' in \emph{IEEE 13th Workshop on
  Positioning, Navigation and Communications (WPNC)}, Oktober 2016. [Online].
  Available: \url{http://elib.dlr.de/108666/}
\BIBentrySTDinterwordspacing

\bibitem{GPUJZ2017}
\BIBentryALTinterwordspacing
C.~Gentner, R.~P{\"o}hlmann, M.~Ulmschneider, T.~Jost, and S.~Zhang,
  ``Positioning using terrestrial multipath signals and inertial sensors,''
  \emph{Mobile Information Systems}, 2017. [Online]. Available:
  \url{http://elib.dlr.de/114476/}
\BIBentrySTDinterwordspacing

\bibitem{PK2011}
Z.~Pi and F.~Khan, ``An introduction to millimeter-wave mobile broadband
  systems,'' \emph{IEEE Communications Magazine}, vol.~49, no.~6, pp. 101--107,
  June 2011.

\bibitem{HSZWM2016}
Y.~Han, Y.~Shen, X.~P. Zhang, M.~Z. Win, and H.~Meng, ``Performance limits and
  geometric properties of array localization,'' \emph{IEEE Transactions on
  Information Theory}, vol.~62, no.~2, pp. 1054--1075, Feb 2016.

\bibitem{GGD2017}
\BIBentryALTinterwordspacing
A.~Guerra, F.~Guidi, and D.~Dardari. (2017) {Single Anchor Localization and
  Orientation Performance Limits using Massive Arrays: MIMO vs. Beamforming}.
  [Online]. Available: \url{{https://arxiv.org/abs/1702.01670}}
\BIBentrySTDinterwordspacing

\bibitem{DS2014}
H.~Deng and A.~Sayeed, ``Mm-wave {MIMO} channel modeling and user localization
  using sparse beamspace signatures,'' in \emph{2014 IEEE 15th International
  Workshop on Signal Processing Advances in Wireless Communications (SPAWC)},
  June 2014, pp. 130--134.

\bibitem{W2007}
H.~Wymeersch, \emph{{Iterative Receiver Design}}.\hskip 1em plus 0.5em minus
  0.4em\relax {Cambridge University Press}, 2007.

\bibitem{SLG2004}
E.~Sharon, S.~Litsyn, and J.~Goldberger, ``An efficient message-passing
  schedule for ldpc decoding,'' in \emph{2004 23rd IEEE Convention of
  Electrical and Electronics Engineers in Israel}, Sept 2004, pp. 223--226.

\bibitem{KTB2011}
D.~Kroese, T.~Taimre, and Z.~Botev, \emph{Handbook of Monte Carlo
  Methods}.\hskip 1em plus 0.5em minus 0.4em\relax Wiley, 2011.

\bibitem{MWB2018b}
\BIBentryALTinterwordspacing
R.~Mendrzik, H.~Wymeersch, and G.~Bauch. (2018) {Joint Localization and Mapping
  through Millimeter Wave MIMO in 5G Systems - Extended Version}. [Online].
  Available: \url{{http://arxiv.org/abs/1804.04417}}
\BIBentrySTDinterwordspacing

\bibitem{WLW2009}
H.~Wymeersch, J.~Lien, and M.~Win, ``{Cooperative Localization in Wireless
  Networks},'' \emph{Proceedings of the IEEE}, vol.~97, 2009.

\end{thebibliography}

\end{document}